\documentclass[%
 aip,
 amsmath,amssymb,
 reprint,%
]{revtex4-1}

\usepackage{graphicx}% Include figure files
\usepackage{dcolumn}% Align table columns on decimal point
\usepackage{bm}% bold math
\usepackage[utf8]{inputenc}
\usepackage[T1]{fontenc}
\usepackage{mathptmx}
\usepackage{etoolbox}
\usepackage{color}
\usepackage{siunitx}

\newcommand{\moy}[1]{\left\langle #1 \right\rangle}

\newcommand{\ex}[1]{\mathrm{e}^{#1}}

\newcommand{\dd}[0]{\mathrm{d}}
\newcommand{\ii}[0]{\mathrm{i}}
\newcommand{\kk}[0]{\boldsymbol{k}}

\newcommand{\FF}[0]{\boldsymbol{F}}
\newcommand{\ee}[0]{\boldsymbol{e}}
\newcommand{\rr}[0]{\boldsymbol{r}}
\newcommand{\RR}[0]{\boldsymbol{R}}

\newcommand{\ff}[0]{\boldsymbol{f}}

\newcommand{\kB}[0]{k_{\mathrm{B}}}

\newcommand{\uu}[0]{\boldsymbol{u}}
\newcommand{\jj}[0]{\boldsymbol{j}}

\newcommand{\ab}[0]{\alpha\beta}

\definecolor{darkblue}{rgb}{0,0,0.6}
\definecolor{darkred}{rgb}{0.6,0,0}
\usepackage[colorlinks=true,urlcolor=darkblue,citecolor=darkblue,linkcolor=darkred]{hyperref}

\makeatletter
\def\@email#1#2{%
 \endgroup
 \patchcmd{\titleblock@produce}
  {\frontmatter@RRAPformat}
  {\frontmatter@RRAPformat{\produce@RRAP{*#1\href{mailto:#2}{#2}}}\frontmatter@RRAPformat}
  {}{}
}%
\makeatother
\begin{document}

%\preprint{AIP/123-QED}

\title[On analytical theories for conductivity and self-diffusion in concentrated electrolytes]{On analytical theories for conductivity and self-diffusion in concentrated electrolytes}
% Force line breaks with \\
\author{Olivier Bernard}
\affiliation{Sorbonne Universit\'e, CNRS, Laboratoire PHENIX (Physicochimie des Electrolytes et Nanosyst\`emes Interfaciaux), 4 place Jussieu, 75005 Paris, France}
\author{Marie Jardat}
\affiliation{Sorbonne Universit\'e, CNRS, Laboratoire PHENIX (Physicochimie des Electrolytes et Nanosyst\`emes Interfaciaux), 4 place Jussieu, 75005 Paris, France}
\author{Benjamin Rotenberg}
\affiliation{Sorbonne Universit\'e, CNRS, Laboratoire PHENIX (Physicochimie des Electrolytes et Nanosyst\`emes Interfaciaux), 4 place Jussieu, 75005 Paris, France}
\author{Pierre Illien}
 \email{pierre.illien@sorbonne-universite.fr}
\affiliation{Sorbonne Universit\'e, CNRS, Laboratoire PHENIX (Physicochimie des Electrolytes et Nanosyst\`emes Interfaciaux), 4 place Jussieu, 75005 Paris, France}

\date{\today}

\begin{abstract}
Describing analytically the transport properties of electrolytes, such as their conductivity or the self-diffusion of the ions, has been a central challenge of chemical physics for almost a century. In recent years, this question has regained some interest in light of Stochastic Density Field Theory (SDFT) -- an analytical framework that allows the approximate determination of density correlations in fluctuating systems. In spite of the success of this theory to describe dilute electrolytes, its extension to concentrated solutions raises a number of technical difficulties, and requires simplified descriptions of the short-range repulsion between the ions. In this article, we discuss recent approximations that were proposed to compute the conductivity of electrolytes, in particular truncations of Coulomb interactions at short distances. We extend them to another observable (the self-diffusion coefficient of the ions) and compare them to earlier analytical approaches, such as the mean spherical approximation and mode-coupling theory. We show how the treatment of hydrodynamic effects in SDFT can be improved, that the choice of the modified Coulomb interactions significantly affects the determination of the properties of the electrolytes, and that comparison with other theories provides a guide to extend SDFT approaches in this context. 
\end{abstract}

\maketitle

\section{Introduction}

Predicting accurately the transport properties of electrolytes, such as their bulk conductivity when submitted to a constant electric field, or the self-diffusion coefficient of the ions, is a central issue of chemical physics, with applications in many research domains, from electrochemistry to soft matter physics. From a theoretical perspective, deriving analytical expressions of these observables starting from elementary principles has been a continuing problem of statistical mechanics since the 1920s. Its difficulty lies in the fact this it intertwines the Brownian diffusion of the ions, their pair interactions (electrostatics at long-range, steric at short-range) and hydrodynamic effects. 

Determining the variation of transport coefficients as a function of electrolyte concentration was initiated by Debye and completed by Onsager\cite{Onsager1927}.   
By considering the ions as point particles in a continuous medium, Onsager's equations led to  limiting laws of evolution of these quantities as the square root of the concentration. 
More realistic descriptions of the solution have provided mathematical expressions that extend the scope of this theory. In particular, starting from the Smoluchowski equations with $2 N$ particles, two-ions densities evolution equations have been derived, which generalize the Onsager equations \cite{Falkenhagen1968,Falkenhagen1971}.  These studies provide validation of the Onsager's equations for applications at low concentrations. Assessing the ionic conductivity involves the calculation of the average ion flux in a uniform steady-state system, in the presence of an external electric field that is sufficiently weak to ensure a linear response.

The case of dense electrolytes is more challenging, since it requires to account for the short-range repulsion between the ions, which becomes predominant when concentration increases.  From an analytical perspective, this requires an 
accurate description of equilibrium pair distribution functions. For ionic concentrations higher than \SI{0.1}{\mol\per\liter}, the replacement of the distribution functions of  Debye and H{\"u}ckel by those given by the hypernetted chains (HNC) or mean spherical approximation (MSA)  theories turned out to be decisive\cite{Bernard1992,Bernard1992a,Chhih1994,Dufreche2005}. 
At last, mode-coupling theory (MCT), combined with time-dependent density functional theory, led to self-consistent expressions of the relaxation and allowed its time dependence to be obtained accurately\cite{Chandra1999,Chandra2000,Dufreche2002,ContrerasAburto2013,Aburto2013}.  

From a computational point of view, in the 1970s, new simulation methods, that relied on continuous or implicit descriptions of the solvent, allowed efficient computations for simple electrolytes \cite{Turq1977}, and accurate descriptions of their properties \cite{Trullas1990,Kunz1990,Canales1998}. Later on, the effect of hydrodynamic interactions was added in these simulation schemes \cite{Jardat1999,Dahirel2013}. These were compared with the aforementioned analytical schemes, and it was shown that, to describe highly charged and concentrated electrolytes, the joint use of Smoluchowski's equations and these equilibrium distributions  together led to a quantitative representation of the simulations \cite{Dahirel2020}. Finally, molecular simulations of electrolytes with an explicit description of the solvent were also employed to predict transport properties successfully \cite{Hoang23,Lesnicki2020,Lesnicki2021}.

Recently, the analytical descriptions of bulk electrolytes has gained a renewed interest in the context of Stochastic Density Field Theory (SDFT). This approach consists in describing the positions of the particles in the suspension as a collection of interacting overdamped Langevin processes. Using It\^o calculus\cite{Gardiner1985}, and relying on the seminal works of Kawasaki\cite{Kawasaki1994} and Dean\cite{Dean1996}, stochastic evolution equations for the density fields of the particles can be derived. Although these equations are quite difficult to study in their original form, linearized expressions were shown to provide accurate estimates of transport coefficients in different kinds of suspensions, made of either charged or neutral particles -- a detailed review of these results will be provided in the next Section. However, in spite of its convenience, this linearization of SDFT must be viewed as a perturbative expansion: it only holds when the interactions between the particles are sufficiently weak, which implies that short-range hardcore repulsion between the ions cannot be accounted for. Recently, Avni \textit{et al.}\cite{Avni2022} proposed a way to circumvent this difficulty by truncating the Coulomb interaction potential below a cut-off radius -- an approach that provided acceptable estimates for the conductivity of simple electrolytes up to concentrations of a few \SI{}{\mol\per\liter}. This linearized SDFT approach also led to an analytical description of the Wien effect \cite{Avni2022a}, namely nonlinear corrections to Ohm's law, that was predicted within Onsager's approach \cite{Fuoss1957,Fuoss1962}, and that also received a surge of interest with explicit- and implicit-solvent simulations \cite{Lesnicki2020,Lesnicki2021}. 

The goal of this article is to discuss the use of  linearized SDFT to study analytically the transport and diffusion properties of concentrated electrolytes. We first give a brief overview of SDFT and recall in which contexts its linearized version has been applied. Then, relying on the modified Coulomb potential introduced  in Refs.~\citenum{Adar2019,Avni2022} that is technically compatible with linearized SDFT, we study systematically the electrostatic and hydrodynamic contributions to the conductivity and show how they differ from earlier theoretical approaches, such as the mean spherical approximation (MSA), or Brownian dynamics simulations, which both include more realistic short-range interactions. From this observation, we further show that the divergence of the hydrodynamic contribution to conductivity, that limits the use of linearized SDFT to moderate concentrations, can be corrected using a refined description of the boundary conditions for the solvent at the surface of the ions. Then, relying on a recent extension of SDFT~\cite{Jardat2022a}, we compute analytically the self-diffusion coefficient of the ions in the electrolyte and compare it to earlier mode-coupling results. Finally, we offer a detailed discussion of the approximate interactions potentials that are compatible with linearized SDFT, and show how their choice strongly affects the outcome of the calculations for both observables (conductivity and self-diffusion).

\section{A brief reminder of SDFT}

\subsection{General equations}

We first recall the main equations of stochastic density field theory (SDFT) and their derivation. For simplicity, we consider a binary mixture of monovalent ions ($N$ cations and $N$ anions of charge $e_\pm=\pm e$), submitted to a constant external field $\boldsymbol{E}_0 = E_0 \ee_z$.
Their respective mobilities will be denoted by $\mu_\pm$, and their bare diffusion coefficients by $D_\pm$ (we assume that the fluctuation-dissipation relation holds, in such a way that $D_\pm=\mu_\pm \kB T$, where $T$ is the temperature of the solution and $k_B$ Boltzmann's constant). The starting point of stochastic density field theory (SDFT) is the set of coupled overdamped Langevin equations obeyed by the positions of the cations and anions $\rr_1^\pm,\dots,\rr_N^\pm$:
\begin{equation}
\label{overdamped_Lan}
\frac{\dd}{\dd t}\rr^\pm_a(t) = \mu_\pm \FF_a^\pm + \uu(\rr^\pm_a(t)) + \sqrt{2D_\pm} \boldsymbol{\xi}_a^\pm(t),
\end{equation}
where $ \boldsymbol{\xi}_a^\pm(t)$ is a unit white noise:
\begin{align}
\label{}
   \langle \xi_{a,i}^\pm (t) \rangle &=0 ,  \\
   \langle \xi_{a,i}^\alpha (t) \xi_{b,j}^\beta (t') \rangle & = \delta_{ab}\delta_{\alpha\beta}\delta_{ij} \delta (t-t'),
\end{align}
where $i$ and $j$ denote Cartesian coordinates, $\uu$ is the velocity field of the solvent, and $\FF_a^\pm$ is the total force undergone by the considered ion $a$, which reads: 
\begin{align}
\label{force_ion}
\FF_a^\pm =& \pm e\boldsymbol{E}_0-\sum_{b\neq a}  \nabla v_{\pm\pm}(\rr^\pm_a(t)-\rr^\pm_b(t))\nonumber\\
& -\sum_{b}  \nabla v_{\mp\pm}(\rr^\pm_a(t)-\rr^\mp_b(t)) .
\end{align}
We denote by   $v_{\alpha\beta}(\rr)$ is the pair interaction potential between one ion of type $\alpha$ and one ion of type $\beta$ separated by a distance $\rr$.

The central ideal of SDFT is to start from the $2N$  overdamped coupled Langevin equations \eqref{overdamped_Lan}, and to derive instead the evolution of the stochastic density fields of cations and anions, defined as
\begin{equation}
\label{def_nalpha}
n_\alpha(\rr,t) = \sum_{a=1}^N \delta(\rr-\rr^\alpha_a(t)).
\end{equation}
Using Ito calculus and following the seminal approaches by Dean \cite{Dean1996} and Kawasaki \cite{Kawasaki1994}, one can show that the densities obey the following stochastic equations 
\begin{equation}
\label{eq_cons}
\partial_t n_\alpha = -\nabla \cdot \jj_\alpha,
\end{equation}
with the fluxes
\begin{equation}
\label{eq_flux}
\jj_\alpha = n_\alpha \uu -D_\alpha \nabla n_\alpha + \mu_\alpha \ff_\alpha - \sqrt{2D_\alpha n_\alpha} \boldsymbol{\zeta}_\alpha,
\end{equation}
where $\boldsymbol{\zeta}_\alpha(\rr,t)$ are space-dependent unit white noises:
\begin{align}
\langle \boldsymbol{\zeta}_\alpha(\rr,t) \rangle &= 0 ,\\
\langle \zeta_{\alpha,i}(\rr,t) \zeta_{\beta,j}(\rr',t') \rangle &= \delta_{\alpha\beta}\delta_{ij}\delta(\rr-\rr')\delta(t-t'), \label{corr_zeta}
\end{align}
and $\ff_\alpha$ is the force density originating from the external electric field $\boldsymbol{E}_0$ and the interactions between the ions:
\begin{equation}
\label{eq_force}
\ff_\alpha = n_\alpha e_\alpha \boldsymbol{E}_0 - n_\alpha \sum_{\beta=\pm} \int \dd \rr'\; n_\beta(\rr') \nabla v_{\alpha\beta}(|\rr-\rr'|).
\end{equation}
The simplest way to account for hydrodynamic effects is to assume that velocity field $\uu$ and {the} pressure field $p$ are the solutions of the following equations: 
\begin{align}
&	\nabla \cdot \uu =0, \\
&	\eta \nabla^2 \uu = \nabla p - \ff_+ - \ff_-,
\end{align}
which correspond respectively to the incompressibility condition and to Stokes equation ($\eta$ denotes the dynamic viscosity of the solvent). {Note that this is not the most general way to account for hydrodynamic interactions \cite{Donev2014},  but that it is valid in the linearized limit \cite{Peraud2017}.}

\subsection{Linearization}

The evolution equations of the density fields $n_\alpha(\rr,t)$ [Eqs. \eqref{eq_cons}, \eqref{eq_flux} and \eqref{eq_force}], usually called `Dean-Kawasaki' (DK) equations, are exact and completely explicit. However, under this form, they are of limited practical use for several reasons: (i) first, the quantities $n_\alpha$, defined as sums of delta-functions [Eq. \eqref{def_nalpha}] are very singular and lack physical interpretation (however, in spite of these singular definitions, the well-posedness and meaning of the DK equations has been the subject of several recent works in the mathematical\cite{Fehrman2019,Konarovskyi2019} and physical\cite{Donev2014} literature); (ii) {the equations obeyed by the fields $n_\alpha$ are  non-linear} (note in particular the non-local couplings through the convolution integrals in Eq. \eqref{eq_force}); (iii) they involve multiplicative noise, as seen in Eq. \eqref{eq_flux}. { Therefore, one needs to resort to further approximations to use these evolution equations.}

For instance, these equations can be linearized, assuming that the density fields remain close to a constant value and that the fluctuations around this base value remain small ($n_\alpha = n_\alpha^0+\delta n_\alpha$ with $\delta n_\alpha \ll n_\alpha^0$). For our simple monovalent electrolyte, we will denote by $n$ the average density of cations and anions: $n=n_+^0=n_-^0$. In the absence of hydrodynamic effects, such a linearisation was first proposed in the particular situation of non-charged colloids interacting via soft potentials \cite{Demery2014}, {and to study the Casimir force between plates in an electrolyte \cite{Dean2014}}. {The validity of this approximation  is restricted to a limited range of parameters.} In spite of this, this linearization allows explicit calculations in different nonequilibrium settings, which made it quite successful over the years. Indeed, it was used  in different contexts: microrheology of colloidal suspensions \cite{Demery2014, Demery2019,Demery2015}, active matter  \cite{Feng2021,Poncet2021b,Tociu2019,Fodor2020,{Tociu2020}, Martin2018}, driven binary mixtures \cite{Poncet2016}.  More recently, it was applied to the study of electrolytes, in order to compute the conductivity of `concentrated' electrolytes without \cite{Demery2015a} and with hydrodynamic interactions \cite{Peraud2016,Donev2019, Peraud2017, Avni2022,{Avni2022a}}, the density-density and charge-charge correlations \cite{Frusawa2020,Frusawa2022},  fluctuation-induced forces between walls \cite{Mahdisoltani2021a,Mahdisoltani2021}, or ionic fluctuations in finite observation volumes \cite{Minh2023}.

\subsection{Modified Coulomb potential}

The linearization of the SDFT equations can be seen as a perturbative approach, that is only valid when the interactions  between  particles are sufficiently `weak'. Moreover, from a computational perspective, the resolution of the linearized equations only holds when the potentials $v_{\alpha\beta}(\rr)$ can be Fourier-transformed. Since this is the case for Coulomb interactions, electrolytes can be studied within linearized SDFT. Indeed, the standard Coulomb potential for monovalent ions $u_{\alpha\beta}(r)=z_\alpha z_\beta e^2/(4\pi\varepsilon_0\varepsilon r)$ has a definite Fourier transform $\tilde u_{\alpha\beta} (k) = z_\alpha z_\beta e^2/(\varepsilon_0\varepsilon k^2)$ (which can be computed by introducing a screening $\exp(-\lambda r)$ that regularizes the integrals and taking $\lambda \to 0$).

However, in order to describe concentrated electrolytes, one has to account for the finite size of the ions and the resulting short-range repulsion, which plays an increasing role on the transport properties of the electrolytes as the ionic density increases. For instance, many theoretical approaches have focused on the `primitive model' of electrolytes, where the ion-ion short-range repulsion is modeled by a pure hardcore repulsion. Although very appealing for analytical developments, this model cannot be treated using {linearized SDFT}, since the infinite repulsion energy below a given cutoff value (which is typically the average of the diameters of the ion) makes its Fourier transform singular.

In order to apply {linearized SDFT}, another cutoff of the Coulomb potential was proposed by Avni et al.\cite{Avni2022} (based on earlier work by Adar et al.\cite{Adar2019}): below a given value $a$, the interaction energy is set to zero:
\begin{equation}
\label{mod_pot_Avni}
v_{\alpha\beta}(r) = z_\alpha z_\beta V_\text{co}(r)=z_\alpha z_\beta\frac{e^2}{4\pi \varepsilon_0 \varepsilon r }\theta(r-a),
\end{equation}
where $\theta$ is the Heaviside function. {As it will become apparent throughout the paper, the outcome of the linearized SDFT equations strongly depend on the choice of the ion-ion interaction potentials. For simplicity, when we will refer to `linearized SDFT', it will be implicit that we use the potential defined in Eq. \eqref{mod_pot_Avni} -- this choice will be discussed extensively in Section \ref{sec_mod_pot}. }

Relying on this modified potential, the conductivity of the electrolyte, defined as 
\begin{equation}
\label{ }
\kappa = \lim_{E_0 \to 0} \frac{\moy{J_x}}{E_0},
\end{equation}
where $J_x$ is the current density along the direction $z$ of the applied field and $\moy{\cdot}$ denotes an ensemble average, can be computed within linearized SDFT. It can be split into different contributions:
\begin{equation}
\label{kappa_generic_decompo}
\kappa = \kappa_0 + \kappa_\text{el} + \kappa_\text{hyd},
\end{equation}
where $\kappa_0 = 2 e^2 \bar \mu n $ is the conductivity at infinite dilution ($\mu = (\mu_+ + \mu_-)/2$ is the mean mobility), also known as the Nernst-Einstein conductivity, and where $\kappa_\text{el} $ (resp. $ \kappa_\text{hyd}$) is the contribution originating from electrostatic (resp. hydrodynamic) effects. Here, we simply recall the expressions for the electrostatic and hydrodynamic contributions obtained with the modified potential given in Eq. \eqref{mod_pot_Avni}\cite{Avni2022}:
\begin{align}
\kappa_\text{el}&= - \frac{1}{3\pi} \frac{\kappa_0 \ell_B}{\lambda_D} \int_0^\infty \dd x \; \frac{x^2 \cos^2x \frac{ax}{\lambda_D}}{x^4 + \frac{3}{2}x^2 \cos\frac{ax}{\lambda_D} + \frac{1}{2}\cos^2\frac{ax}{\lambda_D} },   \label{kappa_el_Avni}\\
\kappa_\text{hyd}&=- \frac{2}{\pi} \frac{\kappa_0 r_s}{\lambda_D} \int_0^\infty \dd x \; \frac{ \cos \frac{ax}{\lambda_D}}{\cos\frac{ax}{\lambda_D} + x^2 }, 
\label{kappa_hyd_Avni}
\end{align}
with the Bjerrum length $\ell_B=e^2/(4\pi\epsilon_0\epsilon \kB T)$ (for numerical evaluations, we will take the value $\ell_B=\SI{7}{\angstrom}$, corresponding to water at \SI{25}{\celsius}) and the Debye screening length $\lambda_D = 1/\sqrt{8\pi \ell_B n}$. Interestingly, we show in Section \ref{app_MSA} that both these expressions can be derived starting from linearized transport equations, completed with approximate expressions of the ion-ion equilibrium distribution functions, that are derived using the expression of the potential given in Eq. \eqref{mod_pot_Avni}, and the Ornstein-Zernike equation.

\section{Fuoss-Onsager approach and Mean Spherical Approximation}
\label{app_MSA}

\subsection{Outline of the calculation}
\label{app_MSA_outline}

In this Section, we present expressions of the hydrodynamic and electrostatic contributions to conductivity ($\kappa_\text{hyd}$ and $\kappa_\text{el}$), whose derivations rely on the theory by Fuoss and Onsager \cite{Onsager1932}.  In this approach, the electric field applied to the electrolyte $\boldsymbol{E}_0=E_0\ee_z$ is assumed to be very small, in such a way that the structure and dynamics of the fluid are   weakly perturbed from equilibrium. Within this linear-response description, and relying on the physical considerations detailed below, the contributions $\kappa_\text{hyd}$ and $\kappa_\text{el}$ are  expressed in terms of the ion-ion distribution functions. The latter  can be evaluated in different ways: using the Mean Spherical Approximation, or using the modified Coulomb potential considered in this paper.

\emph{Mean Spherical Approximation.---} First, the ion-ion distribution functions can be computed within the Mean Spherical Approximation applied to the so-called `primitive model' of electrolytes, where the pair potentials between ions of types $\alpha$ and $\beta$ read:
\begin{equation}
v_{\alpha\beta}(r)=\frac{z_\alpha z_\beta e^2}{4\pi \varepsilon_0\varepsilon r} + v_\mathrm{HS}(r),
\label{primitive_model_pot}
\end{equation}
where  $v_\mathrm{HS}(r)$ denotes the simple hard-sphere potential  (infinite for distances smaller thant the average diameter of the ions, and 0 otherwise). The pair distribution function $g_{\alpha\beta}(r)$ between two ions of respective types $\alpha$ and $\beta$, separated by a distance $r$ and interacting via the potential defined in Eq. \eqref{primitive_model_pot}, cannot be computed exactly for a finite density of ions, but can be approximated. Let us introduce the total pair correlation function $h_{\alpha\beta}(r)=g_{\alpha\beta}(r)-1$, and the direct correlation function $c_{\alpha\gamma} (r) $, defined through the Ornstein-Zernike relations \cite{Hansen1986}
\begin{equation}
h_{\alpha\beta}(r)=c_{\alpha\beta}(r) + n \sum_\gamma  \int \dd \rr' \; c_{\alpha\gamma} (|\rr-\rr'|) h_{\gamma\beta}(r'),
\label{OZ}
\end{equation}
which takes the simpler, deconvoluted form in Fourier space:
\begin{equation}
\tilde h_{\alpha\beta}(k)=\tilde c_{\alpha\beta}(k) + n \sum_\gamma \tilde c_{\alpha\gamma} (k) \tilde h_{\gamma\beta}(k).
\label{OZ_Fourier}
\end{equation}
In the limit $r\to \infty$, the direct correlation function has the exact expression: $c_{\alpha\beta}(r) =-v_{\alpha\beta}(r)/(\kB T)$. The idea of  MSA  is to keep this expression even for finite values of $r$. Note that assuming that it holds for \emph{any} value of $r$ is usually called the `random phase approximation', and is known to be particularly accurate for softcore potentials. However, in the present case, this expression is obviously wrong at very short distances, where ions cannot overlap because of hardcore exclusion. Therefore, the closing relation on which MSA relies is chosen as:
\begin{equation}
\begin{cases}
  g_{\alpha\beta}(r)=0    & \text{if $r<a$}, \\
  c_{\alpha\beta}(r)=-v_{\alpha\beta}(r)/(\kB T)    & \text{if $r>a$}.
\end{cases}
\label{MSA_approx}
\end{equation}
Let us emphasize that the first equation is exact, while the second one is the key hypothesis of MSA.

{The procedure to determine the ion-ion distribution function dates back to the 1970s, and can be summarized as follows. The piecewise approximation given in Eq. \eqref{MSA_approx} is used in Eq. \eqref{OZ} with a procedure due to Baxter \cite{Baxter1970}. This yields a general matrix equation, that was solved explicitly for the primitive model electrolytes by Blum and and Hoye. We refer the reader to Ref. \citenum{Blum1975} for the general method of resolution, and to Ref. \citenum{Blum1977} for expressions of the pair correlation functions. For completeness, we give in Appendix \ref{c_MSA} the expressions of the direct correlation functions $\tilde c_{\alpha\beta}(k) $, from which the ion-ion distribution functions, that are then used as input in the Fuoss-Onsager approach, can be determined. }

\emph{Modified Coulomb potential.---} Second, the ion-ion distribution functions can be determined using the modified potential defined in Eq. \eqref{mod_pot_Avni} {(importantly, in this situation, the distribution functions are not expected to vanish at short distances, as opposed to the key assumption of MSA)}. We then find that the expressions derived from the linearized SDFT approach are retrieved. This is detailed in what follows.

\subsection{Hydrodynamic contribution to the conductivity}

We first evaluate the hydrodynamic contribution to the conductivity. When a given ion is set in motion by an external field $\boldsymbol{E}_0=E_0 \boldsymbol{e}_z$, it drags the solvent, whose velocity field will in turn affect the motion of each ion. In order to estimate this effect, we need to calculate $ \delta \uu_{\alpha}^\text{hyd} (\rr)$, the increment of velocity felt by a given ion of type $\alpha$ located at position $\rr$ under the effect of the electric field. It is the solution of the Stokes equation $\eta \nabla^2 \delta \uu_\alpha^\text{hyd} = \nabla p - \sum_\beta \ff_\beta$ (completed with the incompressibility condition $\nabla \cdot \delta \uu_\alpha^\text{hyd}=0$), where $\ff_\beta$ is the force density due to the displacement of the ions of type $\beta$, which can be written in terms of the distribution functions as:
\begin{equation}
\ff_\beta(\rr) \simeq n z_\beta e h^0_{\alpha \beta}(r) \boldsymbol{E}_0.
\end{equation}
Here, $h^0_{\alpha \beta}(r) $ denotes the  equilibrium distribution function between two ions of types $\alpha$ and $\beta$ separated by a distance $r$. Before evaluating this distribution function, we first write the solution of the Stokes equation, which reads
\begin{equation}
\delta \uu_{\alpha}^\text{hyd} (\rr) =\sum_\beta \int \dd \rr' \;  \mathbf{O}(\rr-\rr') \cdot \ff_\beta(\rr'),
\end{equation}
where $\mathbf{O}(\rr-\rr')$ is the Oseen tensor, recalled in Section \ref{RPhydro}. Writing $\int \dd \rr' \;  \mathbf{O}(\rr-\rr') \cdot \ff_\beta(\rr')= \int \frac{\dd\kk}{(2\pi)^3} \ex{\ii \kk\cdot \rr}\tilde{\mathbf{O}}(\kk) \cdot \tilde{\ff}_\beta(\kk)$, and taking $\rr=\boldsymbol{0}$ with no loss of generality, we get
\begin{equation}
\delta \uu_{\alpha}^\text{hyd}=\sum_\beta \frac{1}{8 \pi^3 \eta} \int \dd \kk \left( \frac{\tilde \ff_\beta}{k^2}-\frac{\kk (\kk \cdot \tilde \ff_\beta)}{k^4}\right),
\end{equation}
where we used the expression of the Fourier transform of the Oseen tensor given in Eq. \eqref{OseenFT}. Performing the angular integrals (using spherical coordinates for $\kk$, with the polar angle measured with respect to the orientation of $\boldsymbol{E}_0$), we find, for the projection of the velocity increment along $z$,
\begin{equation}
\label{velo_increment_Fuoss}
 \delta u_{\alpha,z}^\text{hyd} 
	 = \frac{n}{3 \pi^2 \eta} \sum_{\beta}  e z_{\beta} {E}_0 \int_0^{\infty} dk 
 \; \tilde h_{\alpha \beta}^{0}(k),
\end{equation}  
where $\tilde h_{\alpha \beta}^{0}(k)$ is the Fourier transform of the equilibrium distribution function between ions of type $\alpha$ and $\beta$, which can either be estimated within MSA, or using the modified Coulomb potential.

 \emph{Mean Spherical Approximation.---} If correlation functions from MSA are used, $\delta u_{\alpha}^\text{hyd}$  and $\kappa_\text{hyd}$
  can be evaluated explicitly. It was obtained\cite{Bernard1992}
\begin{equation}
	 \delta u_{\alpha}^\text{hyd,MSA}= - \frac{z_{\alpha}e E_0}{3 \pi \eta} \frac{\Gamma}{ 1 + \Gamma a },
\end{equation}
 where the screening parameter $\Gamma$ is linked to the Debye length $\lambda_D$ by the relation 
\begin{equation}
\Gamma = \frac{\lambda_D^{-1}}{ 1 + \sqrt{ 1 + 2 a/\lambda_D}}.
\end{equation}
Deducing $\kappa_\text{hyd} = \frac{ne}{E_0}\sum_{\alpha}z_\alpha \delta u_{\alpha}^{hyd}$, we find the hydrodynamic contribution to conductivity, normalized by its value at infinite dilution:
\begin{equation}
	 \frac{\kappa_\text{hyd}^\text{MSA}}{\kappa_0}= -  \frac{\Gamma a}{ 1 + \Gamma a }.
  \label{kappa_hyd_MSA}
\end{equation}
This is the expression that was used to plot the hydrodynamic contribution on Fig. \ref{conductivity_contributions}.

 \emph{Modified Coulomb potential.---} When the ions interact via the modified Coulomb potential [Eq. \eqref{mod_pot_Avni}], i.e. when the short-range repulsion is not purely hardcore, the distribution function can be estimated without MSA. 
Using the random phase approximation  $\tilde{c}_{\alpha\beta}(k) =-\tilde{v}_{\alpha\beta}(k)/\kB T$ (i.e. $\tilde{c}_{\alpha\beta}(k) = -4\pi \ell_B z_\alpha z_\beta/k^2$ for the Coulomb potential), the coupled Ornstein-Zernike equations [Eq. \eqref{OZ_Fourier}] can be solved, and one finds 
  \begin{equation}
  \tilde h^\mathrm{Coul}_{\alpha \beta}(k) =\frac{ -4 \pi \ell_{B} z_{\alpha} z_{\beta}/k^2 }{1+2n\cdot 4 \pi \ell_{B}/k^2 } =\frac{ -4 \pi \ell_{B} z_{\alpha} z_{\beta} }{k^2+\lambda_D^{-2} }  ,
\end{equation}
which coincides with the expression  usually obtained from the Debye-H\"uckel approximation\cite{Pitzer1977}.

This prompts us to treat the modified Coulomb potential in a similar way. Under the random phase approximation, and using the expression of the Fourier transform of $v_{\alpha\beta}(r)$, we write the direct correlation function as $\tilde c_{\alpha \beta}(k) = -4 \pi \ell_{B} z_{\alpha} z_{\beta} \cos{(ka)}/k^2$. Solving the Ornstein-Zernike equations yields
 \begin{equation}
 \tilde h_{\alpha \beta}^{0}(k) = -4 \pi \ell_{B} z_{\alpha} z_{\beta} \frac{\cos{(ka)}}
 { k^2 + \lambda_D^{-2} \cos{(ka)} }.
 \label{HADAR} 
 \end{equation}
Reinjecting this formula in the expression of the velocity increment from the Fuoss-Onsager approach [Eq. \eqref{velo_increment_Fuoss}] yields
 \begin{equation}
\delta u_{\alpha,z}^\text{hyd} = - \frac{z_{\alpha} e E_0}{3 \pi^2 \eta \lambda_D} 
\int_0^{\infty} dx \frac{ \cos{\left(\frac{ax}{\lambda_D}\right)}}{ 
  \cos{\left(\frac{ax}{\lambda_D}\right)} + x^2},
 \end{equation}
 where we used the change of variable $x=k\lambda_D$. Writing the corresponding contribution to conductivity as $\kappa_\text{hyd} = \frac{ne}{E_0}\sum_{\alpha}z_\alpha \delta u_{\alpha}^{hyd}$, we find exactly the expression of $\kappa_\text{hyd}$ that was computed from linearized SDFT [Eq. \eqref{kappa_hyd_Avni}]. Finally, note that, when taking the limit $a\to0$ in the expression of the hydrodynamic contribution (both in that from MSA \eqref{kappa_hyd_MSA} and in that from linearized SDFT \eqref{kappa_hyd_Avni}), one retrieves the hydrodynamic contribution in the Debye-H\"uckel-Onsager calculation of conductivity (expression recalled in the caption of Fig. \ref{conductivity_contributions}).

\subsection{Electrostatic contribution to the conductivity}

We now turn to the electrostatic contribution to conductivity, that was denoted by $\kappa_\text{el}$ in the main text. The physical origin of this contribution is as follows: at equilibrium, around a given ion, the ionic atmosphere (with a globally opposite sign) is spherical, and the resulting electrostatic force on the ion is zero on average. However, in a nonequilibrium situation, for instance when an external field is applied, the motion of the ions  perturbs the ionic atmosphere: the deformed ionic distribution results in a nonzero net force on the ion. This force is sometimes called `relaxation force' in the literature.

In order to model this effect, one introduces the two-body time-dependent distribution functions $\mathcal{F}_{\alpha\beta}(\rr_1,\rr_2,t)$, namely the probability to find one ion of species $\alpha$ at position $\rr_1$, and one ion of species $\beta$ at position $\rr_2$, at time $t$. Neglecting three-body effects, these functions obey the continuity equation 
\begin{align}
\label{cont_eq}
\partial_t \mathcal{F}_{\alpha\beta} (\rr_1,\rr_2,t) =& -\nabla_{\rr_1}\cdot [\mathcal{F}_{\alpha\beta} (\rr_1,\rr_2,t) \boldsymbol{V}_{\alpha\beta}(\rr_1,\rr_2,t)] \nonumber \\
&-\nabla_{\rr_2}\cdot [\mathcal{F}_{\beta\alpha} (\rr_2,\rr_1,t) \boldsymbol{V}_{\beta\alpha}(\rr_2,\rr_1,t)],
\end{align}
where $\boldsymbol{V}_{\alpha\beta}(\rr_1,\rr_2,t)$ is the velocity of an ion of type $\alpha$ located at $\rr_1$, when there is an ion of type $\beta$ at $\rr_2$, at time $t$. This velocity is typically evaluated as 
\begin{equation}
\boldsymbol{V}_{\alpha\beta}\simeq \boldsymbol{V}^s_{\alpha}+\mu_\alpha e z_\alpha (-\nabla \psi_\alpha  +\boldsymbol{E}_0)- \kB T \mu_\alpha \nabla \ln \mathcal{F}_{\alpha\beta}.
\end{equation}
The first term is the contribution of the background solvent -- it will be ignored here as the electrostatic and hydrodynamic contributions are considered separately, and the latter was considered in the previous section. The second term is the electrostatic contribution, where $\psi_\alpha$ is the electrostatic potential around the ion  $\alpha$. The last term is the diffusion force. Finally, this set of equations is completed by the Poisson equation obeyed by the electrostatic potential $\psi_\alpha$, which reads
\begin{equation}
     \nabla^2 \psi_{\alpha}(\rr) = -\frac{e}{\varepsilon_o \varepsilon_r} \left[ z_{\alpha} \delta(\rr - \rr_{\alpha}) 
  +n  \sum_{\beta}  z_{\beta}  
 h_{\alpha \beta}(r)  \right].
 \label{Poisson_Fuoss}
\end{equation}

We now focus on the stationary limit of the continuity equation \eqref{cont_eq}, and relate the two-body distribution functions to the total pair correlation function as follows:
\begin{align}
\mathcal{F}_{\alpha\beta}(\rr_1,\rr_2)& \simeq n^2 g_{\alpha\beta}(|\rr_1-\rr_2|) \\
&= n^2[1+h_{\alpha\beta}(|\rr_1-\rr_2|)].
\end{align}
With this rewriting, the continuity equation \eqref{cont_eq}, together with the Poisson equation \eqref{Poisson_Fuoss}, gives a closed set of equations for the electrostatic potentials $\psi_\alpha$ and total pair correlation functions $h_{\alpha\beta}$. These are solved by writing these quantities as the sum of an equilibrium term (with exponent $0$) and a nonequilibrium term (with the `prime' exponent), that originates from the external field $\boldsymbol{E}_0$:
 \begin{align} 
 h_{\alpha\beta} &= h^{0}_{\alpha\beta} + h'_{\alpha\beta}, \\
 \psi_{\alpha} &= \psi^{0}_{\alpha} + \psi'_{\alpha} .
 \end{align}
At leading order in the perturbation (i.e. keeping only terms linear in $h'_{\alpha\beta}$, $ \psi'_{\alpha}$ and $\boldsymbol{E}_0$), the continuity equation becomes
 \begin{align} 
 \nonumber
& \left( \mu_{\alpha} +  \mu_{\beta} \right)  \nabla^2 h^{\prime}_{\beta \alpha}(r) +  \kB T e[ z_{\alpha}  \mu_\alpha \nabla^2 \psi_{\beta}^{\prime}(\rr) 
 -z_{\beta} \mu_\beta \nabla^2 \psi_{\alpha}^{\prime}(\rr) ] \\
 &=  \kB T e \left( z_{\alpha} \mu_\alpha - z_{\beta} \mu_\beta\right) \boldsymbol{E}_0 \cdot \nabla   h^0_{\beta \alpha}(r).
 \label{cont_linearized}
 \end{align}
Relying on the linearity of Poisson equation \eqref{Poisson_Fuoss}, the latter can be split into an equilibrium and a nonequilibrium part:
\begin{align}
\nabla^2 \psi^0_{\alpha}(\rr) &= -\frac{e}{\varepsilon_o \varepsilon_r} \left[ z_{\alpha} \delta(\rr - \rr_{\alpha})   +n  \sum_{\beta}  z_{\beta}   h^0_{\alpha \beta}(r)  \right], \\
\nabla^2 \psi'_{\alpha}(\rr) &= n  \sum_{\beta}  z_{\beta}   h'_{\alpha \beta}(r)  . 
\label{Poisson_noneq}
\end{align}
Knowing the equilibrium distribution function $h_{\alpha\beta}^0$ (for instance through mean spherical approximation, or through the random random phase approximation), Eqs. \eqref{cont_linearized} and \eqref{Poisson_noneq} allow the determination of the nonequilibrium contributions to the electrostatic potentials and the distribution functions. Finally, the electrostatic relaxation force is  estimated as
\begin{equation}
 \delta \boldsymbol{F}_{\alpha} = - \sum_{\beta} n_{\beta} \int_0^{\infty} \nabla \nu_{\alpha \beta}(r) 
\; h^{\prime}_{\beta \alpha}(\rr) \; \dd \rr,
\end{equation}
from which one deduces the electrostatic contribution to conductivity:
 \begin{equation}
	 \frac{\kappa_\text{el}}{\kappa_0}=\frac{\delta {F}_{\alpha,z}}{z_{\alpha} e {E}_0} .
\end{equation}

\emph{Mean Spherical Approximation.---} Using the equilibrium distribution functions $h_{\alpha\beta}^0$ computed within MSA, it was found \cite{Bernard1992}
 \begin{align}
	 &\frac{\kappa_\text{el}}{\kappa_0}=\frac{\delta {F}_{\alpha}^{\text{MSA}}}{z_{\alpha} e {E}_0} =  - 
	 \frac{\ell_B  }{6 a \left( 1 + \Gamma a \right)^2} \nonumber \\ 
 &\times \frac{ 1 - e^{- \sqrt{2} a/\lambda_D} }{1 + 2 \sqrt{2} \Gamma \lambda_D + 4 \Gamma^2 \lambda_D^2 
	 \left( 1 - e^{-a/\sqrt{2} \lambda_D} \right) } .
  \label{kappa_el_MSA}
\end{align}
This is the expression plotted on Fig. \ref{conductivity_contributions}.

\emph{Modified Coulomb potential.---} As an alternative, by integration of the Poisson equation, the electric field can also be evaluated  from the convolution product of the Coulomb potential with the surrounding charge distribution. 
{Now, in order to recover the results by Avni et al.\cite{Avni2022},}  the Coulomb potential can be replaced again by the    modified Coulomb potential given in Eq. \eqref{mod_pot_Avni}. 
  The equilibrium contribution to the electrostatic potential is given by 
  \begin{equation}
  \tilde \psi_{\alpha}^0(k) = \frac{4 \pi e}{\varepsilon_0 \varepsilon_r} \frac{\cos{ka}}{k^2} \left[ z_{\alpha} 
   + n\sum_{\beta}  z_{\beta} 
 \tilde h^0_{\alpha \beta}(k) \right] .
  \end{equation}
 Using the relation from the Debye-H\"uckel approximation: $\tilde h^0_{\alpha \beta}(k) = - z_{\alpha} e \tilde \psi_{\beta}^0 /k_B T$, the expression previously found   [Eq. \eqref{HADAR}]  is recovered. 
  The contribution due to the external field is given by 
  \begin{equation}
 \tilde \psi_{\alpha}^{\prime}(\boldsymbol{k}) = 
 \frac{4 \pi e}{\varepsilon_0 \varepsilon_r}  \frac{\cos{ka}}{k^2} \sum_{\beta} n z_{\beta} 
 \tilde h^{\prime}_{\alpha \beta}(\boldsymbol{k}) .
  \end{equation}
 Then, taking the Fourier transform of the continuity equation \eqref{cont_linearized}, and   expressing the electrostatic potentials 
 as a function of $\tilde h^{\prime}_{\alpha \beta}(k)$, 
  the latter can be determined  from $\tilde h^0_{\alpha \beta}(k)$.
  The relaxation force is computed in Fourier space using the relation  $
 \left( \nabla \psi^{\prime}_{\alpha} \right)_{r=0} = - \frac{1}{8 \pi^3} \int d \boldsymbol{k} \; 
  i \boldsymbol{k} \; \tilde\psi^{\prime}_{\alpha}(\boldsymbol{k}) $. This leads to 
  \begin{equation}
\frac{\kappa_\text{el}}{\kappa_0}= - \frac{ \ell_B }
{ 3 \pi \lambda_D} \int_0^{\infty} dx \frac{ x^2 \; \cos^2{\frac{a x}{\lambda_D}} }
{ \left( x^2 + \frac{1}{2} \cos{\frac{a x}{\lambda_D}} \right) \left( x^2 + 
\cos{\frac{a x}{\lambda_D}} \right) } ,
  \end{equation}
  which is exactly the expression of $\kappa_\text{el}$ that was computed from linearized SDFT [Eq. \eqref{kappa_el_Avni}]. Finally, in the limit $a\to0$, both Eqs. \eqref{kappa_el_MSA} and \eqref{kappa_el_Avni} yield the electrostatic contribution to conductivity predicted within the Debye-H\"uckel-Onsager theory (expression recalled in the caption of Fig. \ref{conductivity_contributions}).

%}

\section{Conductivity: Comparing results from linearized SDFT and MSA}

\begin{figure*}
\begin{center}
\includegraphics[width=2\columnwidth]{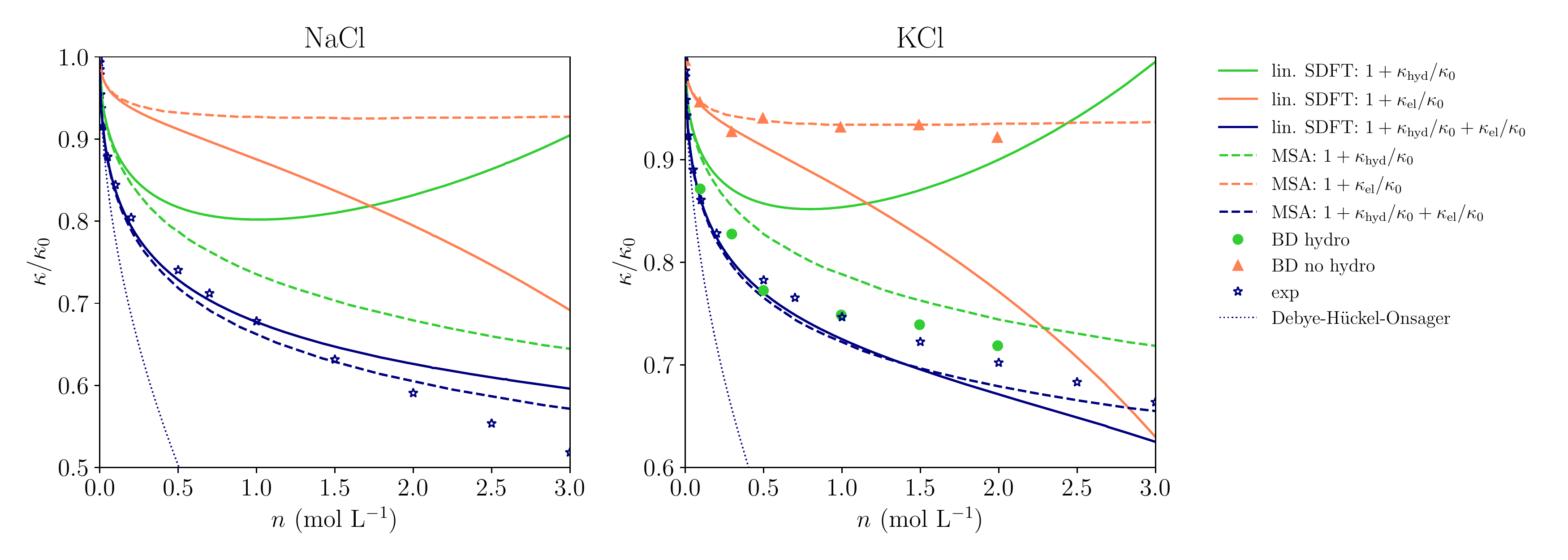}
\caption{\textbf{Comparison between linearized SDFT and MSA.} Conductivity of NaCl (left) and KCl (right) electrolytes, rescaled by their value at infinite dilution $\kappa_0$, obtained from linearized SDFT (solid lines) and MSA  (dashed lines). For KCl, we show results from Brownian dynamics simulations \cite{Jardat1999} (symbols), with and without hydrodynamics. The value of the conductivity is decomposed as the sum of its value at infinite dilution, an electrostatic contribution and a hydrodynamic contribution [Eq. \eqref{kappa_generic_decompo}]. {For both electrolytes, we also show results from experimental measurements of the conductivity (stars)\cite{Miller1966}}. Finally, we represent the classical result from the Debye-H\"uckel-Onsager approach\cite{Onsager1927} (dotted lines), which reads: $\frac{\kappa}{\kappa_0}=1-\left[\frac{\sqrt{2}}{3\sqrt{\pi}}\frac{\ell_B^{1/2}}{\eta\bar \mu} + \frac{2\sqrt{\pi}(\sqrt{2}-1)}{3}\ell_B^{3/2}\right] n^{1/2}$.}
\label{conductivity_contributions}
\end{center}
\end{figure*}

The electrostatic and hydrodynamic contributions are plotted on Fig. \ref{conductivity_contributions} for two simple monovalent electrolytes: NaCl and KCl. As noted in Ref.~\citenum{Avni2022a},  the electrostatic and hydrodynamic contributions to conductivity diverge (to $-\infty$ and $+\infty$, respectively) when the concentration of the electrolyte is too high. In order to discuss the accuracy of the linearized SDFT approach, we compare the behaviors of these two contributions with  earlier analytical results, that relied on the Mean Spherical Approximation (MSA).

\begin{table}[b]
\centering
\begin{tabular}{c c c c c}
\hline
electrolyte & $D_0^+$ ($10^{-9}\text{m}^{2}\cdot\text{s}^{-1}$) & $D_0^-$ ($10^{-9}\text{m}^2\cdot\text{s}^{-1}$) &  $a$ (\AA) &  $r_s$ (\AA)\\
\hline
LiCl & $1.03$ & $2.03$ & $2.57$ & --\\
NaCl & $1.33$ & $2.03$ & $2.83$ & $1.45$\\
KCl & $1.96$ & $2.03$ & $3.19$ & --\\
\hline
\end{tabular}
\caption{Summary of parameter values for the different electrolytes considered in the manuscript.}
\label{numerical_params}
\end{table}

On Fig. \ref{conductivity_contributions}, we plot the conductivity obtained from linearized SDFT [Eqs. \eqref{kappa_generic_decompo}--\eqref{kappa_hyd_Avni}, see Table \ref{numerical_params} for the numerical parameters] and the conductivity obtained from MSA (the underlying assumptions and the derivation of the expressions of $\kappa_\text{el}$ and $\kappa_\text{hyd}$  are given in Section \ref{app_MSA}). For both theories, we plot separately the different contributions, and several comments follow: (i) the main comment is that, although the \emph{total} expression of the conductivity obtained from MSA and linearized SDFT appear to give very similar numerical values, the behavior of their hydrodynamic and electrostatic \emph{contributions} are qualitatively and quantitatively different. In particular, the hydrodynamic contribution predicted by SDFT appear to be non-monotonic. This observation is not in agreement with Brownian dynamics simulations, which can be performed with and without hydrodynamic interactions, in order to decipher the role of the electrostatic and hydrodynamic contributions (simulation results obtained in the case of KCl electrolyte are also shown on Fig. \ref{conductivity_contributions}). In addition, the electrostatic contribution is much more important in linearized SDFT than in MSA, but both effects seem to compensate; {(ii) when compared to experimental measurements, the full expression obtained from linearized SDFT  (blue solid lines)  does not improve upon the MSA prediction (blue dashed lines).} This observation prompted us to look for different ways to improve linearized SDFT in this context. In the next Section, we will see how the treatment of hydrodynamic interactions can be refined. At the end of the manuscript, we will discuss the role played by the details of the modified potential on the outcome of the linearized SDFT calculations.

\section{Accounting for the finite size of the ions in the hydrodynamic equations}
\label{RPhydro}

The result for the hydrodynamic contribution to the conductivity, given in Eq. \eqref{kappa_hyd_Avni}, is obtained by assuming that, from the point of view of hydrodynamics, the ions are point-like, so that their finite size only plays a role in the modified electrostatic potential. Indeed, denoting by $\delta\uu$ and $\delta \rho = \delta n_+ - \delta n_-$  the small perturbations around homogeneous states, the hydrodynamics equations read
\begin{align}
\eta \nabla^2 \delta \uu &= \nabla \delta p -e \delta \rho \boldsymbol{E}_0, \label{hydro1}\\
\nabla\cdot \delta \uu &=0.\label{hydro2}
\end{align}
In the simplest possible description, their solutions are found assuming that the ions are pointlike, and neglecting possible boundary conditions at their surface. From a technical point of view, this is achieved by using the Oseen tensor, defined as
\begin{equation}
O_{ij}(\rr) = \frac{1}{8\pi\eta} \left( \frac{\delta_{ij}}{r}+\frac{r_i r_j}{r^3}\right),
\end{equation}
which is the Green's function to Eqs. \eqref{hydro1} and \eqref{hydro2}  under these assumptions. The  Fourier transform of this tensor reads \footnote{Throughout the paper, we will use the following convention for Fourier transformation:
\begin{align*}
\tilde{f}(\kk) &= \int \dd\rr \; \ex{-\ii \kk\cdot \rr} f(\rr), \\
f(\rr)&= \int \frac{\dd\kk}{(2\pi)^d} \; \ex{\ii \kk\cdot \rr} \tilde{f}(\kk).
\end{align*} }
\begin{equation}
\label{OseenFT}
\tilde O_{ij}(\kk) = \frac{1}{\eta k^2} \left( \delta_{ij}-\frac{k_i k_j}{k^2}\right),
\end{equation}
and the solution of Eqs. \eqref{hydro1} and \eqref{hydro2} is\cite{Avni2022}:
\begin{align}
\delta \tilde \uu & = e \tilde{\mathbf{O}}\cdot \boldsymbol{E}_0 \delta\tilde\rho(\kk), \\
&=  \frac{e E_0}{\eta k^2} \left( \delta_{ij}-\frac{ k_x^2}{k^2}\right) \delta\tilde\rho(\kk).
\end{align}
The expression of the hydrodynamic contribution to the conductivity, given in Eq. \eqref{kappa_hyd_Avni}, follows from this expression.

\begin{figure}
\begin{center}
\includegraphics[width=\columnwidth]{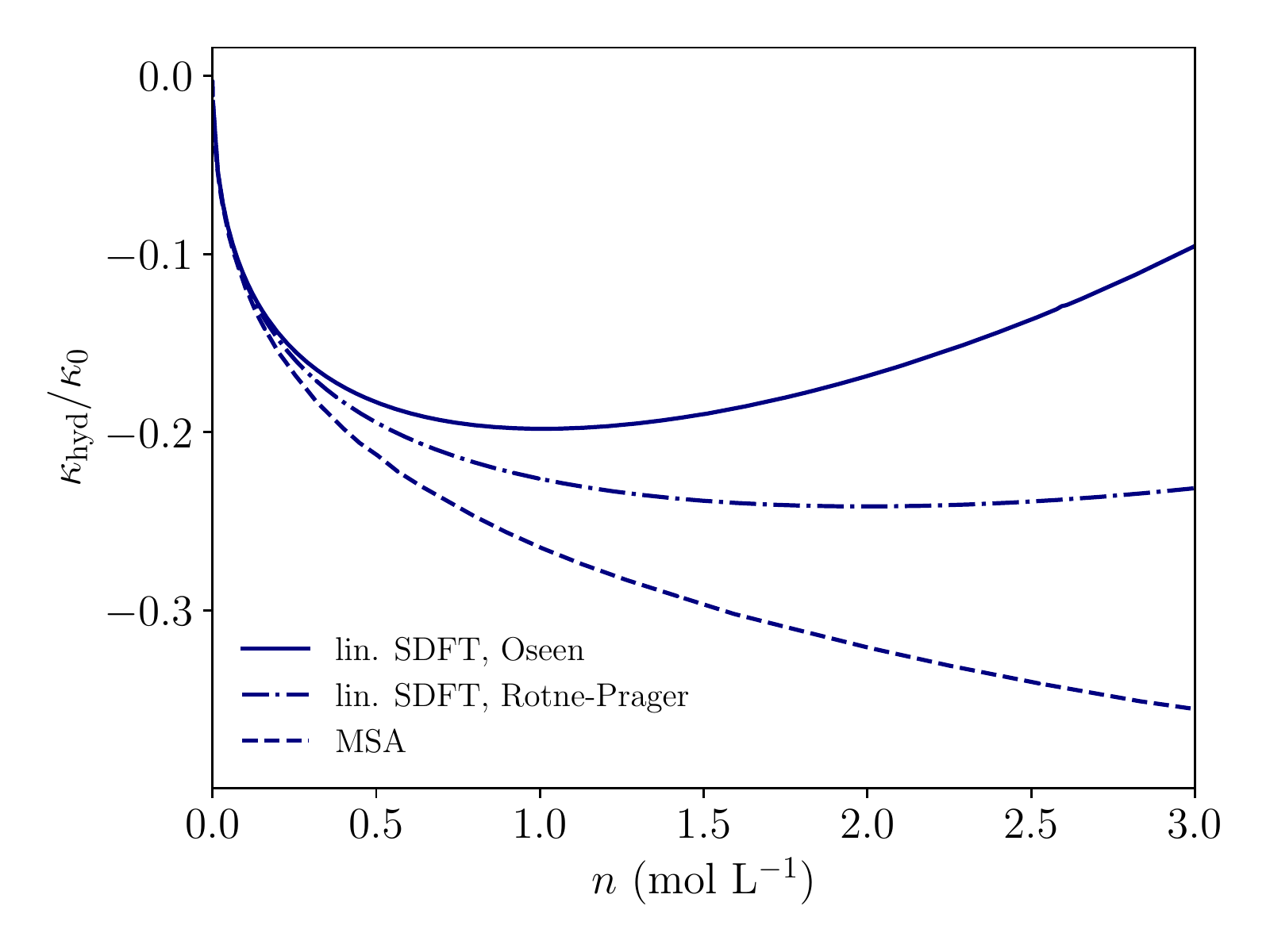}
\caption{\textbf{Hydrodynamic description at the Rotne-Prager level.} Comparison of the different hydrodynamic contributions to the conductivity of NaCl electrolyte, rescaled by their value at infinite dilution $\kappa_0$, obtained from SDFT with the Rotne-Prager tensor (solid lines, Eq. \eqref{kappa_hyd_Rotne}), linearized SDFT with the Oseen tensor (dash-dotted lines, Eq. \eqref{kappa_hyd_Avni}) and MSA (dashed lines, see Section \ref{app_MSA} for details).}
\label{Rotne}
\end{center}
\end{figure}

This calculation can be improved by assuming that, from the point of view of hydrodynamics, the ions actually have a finite radius $r_s$ (the Stokes radius), and that the solvent has a no-slip boundary condition at the surface of the ions: $\uu(r=r_s)=0$. Under these assumptions, the Green's function of Eqs. \eqref{hydro1} and \eqref{hydro2}  is now the Rotne-Prager tensor, whose expression reads, in real space\cite{Kim2005} 
\begin{align}
R_{ij} (\rr) =& \frac{1}{6\pi\eta r_s} \left\{ \frac{3}{4} \left( \frac{r_s}{r}\right) \left( \frac{\delta_{ij}}{r}+\frac{r_i r_j}{r^3}\right)  \right.\nonumber \\
&\left.+ \frac{1}{2} \left( \frac{r_s}{r}\right)^3 \left( {\delta_{ij}}-3\frac{r_i r_j}{r^2}\right)\right\}.
\end{align}
In principle, the anions and the cations have different hydrodynamic radii, so that the expression of the Rotne-Prager tensor would actually depend on the considered pair. However, we assume for simplicity that there is only one radius involved in the Rotne-Prager tensor, which is a reduced Stokes radius $r_s=1/(6\pi\eta\bar \mu)$ (see Table \ref{numerical_params} for numerical values). Note that this keeps the hydrodynamic description consistent with the modified potential given in Eq. \eqref{mod_pot_Avni}, in which one assumes that the cutoff radius $a$ is the same for all the ion pairs. The Fourier transform of the Rotne-Prager tensor can be deduced quite easily from that of the Oseen tensor by noting that \cite{Kim2005}:
\begin{equation}
\nabla^2  \left( \frac{\delta_{ij}}{r}+\frac{r_i r_j}{r^3}\right) = 2  \left( \frac{\delta_{ij}}{r^3}-3\frac{r_i r_j}{r^5}\right),
\end{equation}
which yields \cite{Beenakker1986}
\begin{equation}
\tilde R_{ij} (\kk) = \frac{1}{\eta}\left( \delta_{ij}-\frac{k_i k_j}{k^2}\right) \left(\frac{1}{k^2} -\frac{r_s^2}{3}\right).
\end{equation}
The expression of the perturbation to the velocity field induced by the displacement of the ions is now $\delta \tilde \uu  = e \tilde{\mathbf{R}}\cdot \boldsymbol{E}_0 \delta\tilde\rho(\kk)$, and the result from $\kappa_\text{hyd}$ now reads:
\begin{align}
\kappa_\text{hyd} = &- \frac{2}{\pi} \frac{\kappa_0 r_s}{\lambda_D}   \int_0^{\frac{2\pi\lambda_D}{r_s}} \dd x \left( 1-\frac{r_s^2 x^2}{3\lambda_D^2}\right) \frac{ \cos \frac{ax}{\lambda_D}}{\cos\frac{ax}{\lambda_D} + x^2 }.
\label{kappa_hyd_Rotne}
\end{align}
Note that introducing the Rotne-Prager tensor makes the $x$-integral (i.e. the $k$-integral since we use the change of variable $x=k\lambda_D$) divergent at large $x$. This is regularized by introducing an upper cutoff which corresponds to the size of the ion. We show on Fig. \ref{Rotne} the linearized SDFT theory corrected by using the Rotne-Prager tensor instead of the Oseen tensor, for NaCl electrolyte. 
Interestingly, it seems like introducing this no-slip boundary condition for the solvent at $r=r_s$ regularizes the hydrodynamic contribution, or at least shifts its divergence to higher concentrations. At this point, it would be tempting to account for higher-order corrections in the treatment of the hydrodynamic interactions. Indeed, the multi-body hydrodynamic tensors, that allows one to go beyond the simple Rotne-Prager treatment and to account for multi-body hydrodynamic interactions, are known analytically  \cite{Kim2005}. However, this would be inconsistent with the treatment of the other interactions between the ions, which are only accounted for at the pair level in the usual DK framework [Eq. \eqref{overdamped_Lan}].

% {\color{red}

% commentaire sur le sens de tout ça

% dans l'espace reel on a la mm chose (cf ...)

% dans l'espace de Fourier, plus complique car les modes sont melanges 

% necessite d'introduire des cutoff à grand $k$

% }

\section{Self-diffusion coefficient}
\label{sec_self_diff}

\subsection{General equations}
\label{sec_general_eqs}

We now turn to another observable, namely the self-diffusion coefficient of a tagged ion in the electrolyte. To compute this quantity, it is not necessary to assume that the electrolyte is submitted to a constant electric field force as in the calculation of the conductivity. Without loss of generality, our goal is to calculate the self-diffusion coefficient of the first cation and denote its position by $\RR(t)\equiv \rr_1^+(t)$. The positions of the ions  are still assumed to obey the overdamped Langevin equations given by Eqs. \eqref{overdamped_Lan}--\eqref{force_ion} (with $\boldsymbol{E}_0=0)$.
We now define the densities 
\begin{align}
\label{}
   n'_+(\rr,t)  &=\sum_{a=2}^N \delta(\rr-\rr_a^+(t)) ,  \\
   n'_-(\rr,t)  &=\sum_{a=1}^N \delta(\rr-\rr_a^-(t)) .
\end{align}
Note that these quantities slightly differ from the densities defined earlier, as the contribution from the ion that plays the role of a tracer has been extracted from the summation. In what follows, we drop the `primes' to ease the notation. The equation of motion for a cation (other than the tracer) then reads
\begin{align}
\label{ }
&\frac{\dd}{\dd t}\rr^+_a(t) =  \uu(\rr_a^+(t)) + \sqrt{2D_+} \boldsymbol{\eta}_a^+(t)\nonumber\\
&-\mu_+\nabla \int \dd\rr' \; V_{\text{co}}(\rr_a^+(t)-\rr')[n_+ -n_- + \delta(\rr'-\RR(t))] ,
\end{align}
and that of an anion reads
\begin{align}
\label{ }
&\frac{\dd}{\dd t}\rr^-_a(t) = \uu(\rr_a^-(t)) + \sqrt{2D_-} \boldsymbol{\eta}_a^-(t)\nonumber\\
&-\mu_-\nabla \int \dd\rr' \; V_{\text{co}}(\rr_a^-(t)-\rr')[-n_+ +n_- - \delta(\rr'-\RR(t))] ,
\end{align}
where we denote by $V_{\text{co}}$ the modified Coulomb potential:
\begin{equation}
\label{mod_pot_Avni_bis}
V_{\text{co}}(r) = \frac{e^2}{4\pi \varepsilon_0 \varepsilon r }\theta(r-a).
\end{equation}

We define $\rho = n_+ - n_-$, and we summarize below the main equations that constitute the starting point of our calculation. The evolution equations for the densities  $n_+$ and $n_-$ are obtained as before with Ito calculation \cite{Dean1996}
	\begin{equation}
	\partial_t n_\alpha + \nabla \cdot \jj_\alpha=0,
	\end{equation}
	with the fluxes:
	\begin{align}
&	 \jj_\alpha= n_\alpha \uu - D_\alpha \nabla n_\alpha - \sqrt{2D_\alpha n_\alpha}\boldsymbol{\zeta}_\alpha \nonumber\\
&	 - \mu_\alpha z_\alpha n_\alpha \int \dd\rr'\; \nabla V_{\text{co}}(\rr-\rr')[\rho(\rr',t)+\delta(\rr'-\RR(t))],
	\end{align}
where $z_\pm=\pm1$, and where the noises $\boldsymbol{\zeta}_\alpha $ have zero average, and correlations given by Eq. \eqref{corr_zeta}. The equation of motion of the tracer particle reads
		\begin{align}
	&\frac{\dd}{\dd t} \RR(t) =  \uu(\RR(t)) +\sqrt{2D_+}\boldsymbol{\eta}_0^+(t) \nonumber\\
	&-\mu_+ \nabla \int\dd \rr'\; V_{\text{co}}(\RR(t)-\rr') \rho(\rr',t) ,
	\end{align}
with $\langle\boldsymbol{\eta}_0^+(t)  \rangle = 0 $ and $\langle{\eta}_{0,i}^+(t) {\eta}_{0,j}^+(t') \rangle = \delta_{ij}\delta(t-t') $. The velocity field of the solvent, denoted by $\uu$, obeys the following hydrodynamic equations (Stokes equation and incompressibility condition):
	\begin{align}
&	\nabla \cdot \uu =0 \\
&	\eta \nabla^2 \uu = \nabla p - \delta(\rr-\RR(t)) \int \dd \rr' \nabla V_{\text{co}} (\rr-\rr') \rho(\rr',t) \nonumber\\
&	- \rho(\rr,t) \int \dd \rr' \; \nabla V_{\text{co}} (\rr-\rr') [\rho(\rr',t)+\delta(\rr'-\RR(t))]
	\end{align}
	where $p$ is the pressure field in the solvent.

These equations are then linearized using the perturbative expansions:
\begin{align}
n_\pm &= n +\delta n_\pm, \\
\rho &= n_+ - n_- = \delta n_+ - \delta n_-, \\
\uu &= \delta \uu, \\
p &= p_0+\delta p.
\end{align}
Note that, although the densities $n_+$ and $n_-$ do not correspond to the exact same number of ions ($N-1$ for the cation density and $N$ for the anion density), we still linearize them around the same constant value $n = N/V$, which is a valid approximation in the thermodynamic limit ($N\to\infty$ and $V\to \infty$ with fixed number density $n=N/V$).
At leading order in these perturbations, we get the following evolution equations for the densities $\delta n_+$ and $\delta n_-$ in Fourier space
		\begin{align}
&\partial_t \delta \tilde n_\alpha = -D_\alpha k^2 \delta \tilde n_\alpha + \sqrt{2D_\alpha n} \kk \cdot \tilde{\boldsymbol{\zeta}}_\alpha(\kk) \nonumber\\
&- \mu_\alpha n z_\alpha k^2 \tilde V_{\text{co}}(\kk) \tilde \rho -\mu_\alpha n z_\alpha k^2 \tilde V_{\text{co}}(\kk) \ex{-\ii \kk \cdot \RR(t)}.
	\end{align}

Interestingly, we observe that the velocity field does not appear anymore in the equations for the density fields $\delta \tilde c_\alpha$. This is due to the incompressibility condition: 
  \begin{equation}
\label{ }
\nabla \cdot (n_\alpha \uu) = \nabla \cdot[(n+\delta n_\alpha)\delta\uu] \simeq n \nabla \cdot \delta \uu = 0 
\end{equation}
The hydrodynamic equations now read:
		\begin{align}
	\nabla \cdot \uu &=0 \\
	\eta k^2 \delta \tilde{\uu}&= -\ii \kk \delta \tilde p 
		\end{align}
	At leading order, the forces exerted by the ions on the solvent vanish, and the effect of solvent flows on the dynamics of the ions is then negligible. Indeed, the interaction terms in the Stokes equation vanish because they either involve contributions of order $\delta \rho^2$ or terms of order $\delta \rho \times \delta_{\RR(t)}$, which are negligible since $| \delta_{\RR(t)}| \sim 1/V$ where $V$ is the volume of the system. Therefore, the choice of the right description for boundary conditions at the surface of the ions, that was discussed in Section \ref{RPhydro}, is here irrelevant .

We finally end up with the set of equations (with $\alpha=\pm1$):
\begin{align}
\label{}
&\frac{\dd}{\dd t}\RR(t)     = + \sqrt{2D_+}\boldsymbol{\eta}_0^+(t)\nonumber\\
&-\mu_+ \nabla \int \dd \rr' V_{\text{co}}(\RR(t)-\rr') [\delta n_+(\rr',t)-\delta n_-(\rr',t)] ,   \label{tracer_mapping}\\
&\partial_t \delta \tilde n_\alpha = -D_\alpha k^2 \delta \tilde n_\alpha + \sqrt{2D_\alpha n} \kk \cdot \tilde{\boldsymbol{\zeta}}_\alpha(\kk)  \nonumber\\
&- \mu_\alpha n z_\alpha k^2 \tilde V_{\text{co}}(\kk)(  \delta \tilde n_+ -  \delta\tilde  n_-) -\mu_\alpha n z_\alpha k^2 \tilde V_{\text{co}}(\kk) \ex{-\ii \kk \cdot \RR(t)}, \label{field_mapping}
\end{align}
that fully determine the dynamics of the tagged ion and that of the ionic density fields within the linearized approximation. The counterpart of these equations when the tracer is an anion can be obtained straightforwardly.

\subsection{Path-integral formulation}
\label{sec:path_integral}

The first equation [Eq. \eqref{tracer_mapping}] can be seen as a simple Langevin equation that describes the motion of the tracer particle.  However, it is not easy to compute the associated effective diffusion coefficient. Indeed, this equation of motion involves on its rhs a functional of the density fields $\delta n_\alpha$, whose evolution equations [Eq. \eqref{field_mapping}] also depends on the tracer position $\RR(t)$. We then end up with a set of equations which are coupled nonlinearly.

A way to treat this coupling was proposed by Dean and D\'emery   \cite{Demery2011}, who proposed a perturbative path-integral method, to compute the effective diffusion coefficient of the tracer at leading order in the coupling between the tracer and a fluctuating environment. Here, the situation is more complicated, because the tracer interacts with two fields at the same time (the density field of cations and that of anions). We recently extended the Dean-D\'emery method to compute perturbatively the diffusion coefficient of a tracer coupled to multiple fields \cite{Jardat2022a}.

We apply this method to the present case and obtain the following expression for the diffusion coefficient of the tagged cation rescaled by its bare value $D_+$ (the formula below applies to both situations where the tracer is a cation or an anion):
\begin{equation}
\label{ }
\frac{D_{\text{eff},+}}{D_+} = 1- \sum_{\alpha,\beta=\pm} \frac{\overline{D}_{\alpha\beta}}{D_+}
\end{equation}
with
\begin{align}
\label{full_exp_D}
&\frac{\overline{D}_{\alpha\beta}}{D_+}= \frac{\mu_+\mu_\beta}{d} \int \frac{\dd^d \kk}{(2\pi)^d} k^4 n z_\alpha \tilde V_{\text{co}}^2 \sum_\gamma z_\gamma \nonumber\\
&\times \sum_{\nu = \pm1}  \frac{2c_{\alpha\beta}^{(\nu)}}{(D_+ k^2+\Lambda_\nu)^2} \left[  \delta_{\gamma\beta} + (D_+ k^2-\Lambda_\nu) \sum_{\epsilon=\pm1} \frac{c_{\gamma\beta}^{(\epsilon)}}{\Lambda_\nu+\Lambda_\epsilon} \right]
\end{align}
where we defined the matrices,
\begin{align}
\label{c_matrices}
\boldsymbol{m } &=
k^2
\begin{pmatrix}
\mu_+(\kB T+n \tilde V_{\text{co}})
 &-\mu_+n \tilde V_{\text{co}} \\
-\mu_- n \tilde V_{\text{co}} 
   &\mu_-(\kB T+n \tilde V_{\text{co}})
\end{pmatrix},\\
\boldsymbol{c^{(\pm)} } &=
\frac{1}{2s}
\begin{pmatrix}
{\pm m_{++}\mp m_{--}+s}
 &\pm 2{m_{+-}} \\
\pm 2{m_{-+}}
   & \mp m_{++}\pm m_{--}+s
\end{pmatrix},
\end{align}
the eigenvalues 
\begin{equation}
\Lambda_\pm =  \frac{m_{++}+m_{--}}{2} \pm \frac{1}{2}\sqrt{(m_{++}-m_{--})^2+4m_{+-}m_{-+}},
\end{equation}
and the quantity
\begin{equation}
s = \sqrt{(m_{++}-m_{--})^2+4m_{+-}m_{-+}}.
\end{equation}
In order to get a more compact expression of the diffusion coefficient, it is convenient to introduce a dimensionless interaction potential $\tilde U(k)= n \tilde{V}_\mathrm{co}(k)/\kB T$. Relying on the fact that the integrand in the $\kk$-integral from Eq. \eqref{full_exp_D} is spherically symmetric to perform the angular integration, specifying this expression to the case $d=3$, we get the following formula, which applies to both situations where the tracer is a cation or an anion ($D_0$ denotes the corresponding bare diffusion coefficient):
\begin{align}
&\frac{D_\text{eff}}{D_0} = 1- \int_0^\infty \dd k\frac{32 k^2}{\pi^2 n} (3+\Delta)(3\Delta \tilde U+2\Delta+\tilde U+2)\Delta \tilde U^2  \nonumber \\
& \times [((\tilde U+3)+(\tilde U+1)\Delta)^2 -S^2]^{-2}  [((\tilde U+1)^2(\Delta+1)^2 -S^2]^{-1} \nonumber \\
\label{Deff_full_exp}
\end{align}
where we introduced the shorthand notation
\begin{equation}
S = \sqrt{(\Delta+1)^2 \tilde U^2 +(\Delta-1)^2(2 \tilde U+1)},
\end{equation}
and where $\Delta = D_0^-/D_0^+$ (resp. $D_0^+/D_0^-$) if the tracer is a cation (resp. an anion). Eq. \eqref{Deff_full_exp} provides an explicit and general expression for the effective diffusion coefficient of a tagged ion in the electrolyte, and it is the main result of this Section. Although the $k$-integral can be divergent depending on the analytical expression of the interaction potential \cite{Demery2011,Jardat2022a}, we will focus on the truncated potential given in Eq. \eqref{mod_pot_Avni}, which does not cause any small-$k$ or large-$k$ divergence, and which yields the following rescaled potential $\tilde u$:
\begin{equation}
\label{ }
\tilde U(k)=  4\pi n\ell_B  \frac{\cos (ka)}{k^2}.
\end{equation}
Consequently, the general expression given in Eq. \eqref{Deff_full_exp} only depends on three parameters: the electrolyte concentration $n$, the cutoff of the truncated potential $a$ (which will typically be measured in units of the Bjerrum length, setting the dimensionless parameter $\bar{a} = a/\ell_B$), and the ratio between the bare diffusion coefficient $\Delta$. We now consider a few limit cases of Eq. \eqref{Deff_full_exp}.

\begin{figure}
\begin{center}
\includegraphics[width=\columnwidth]{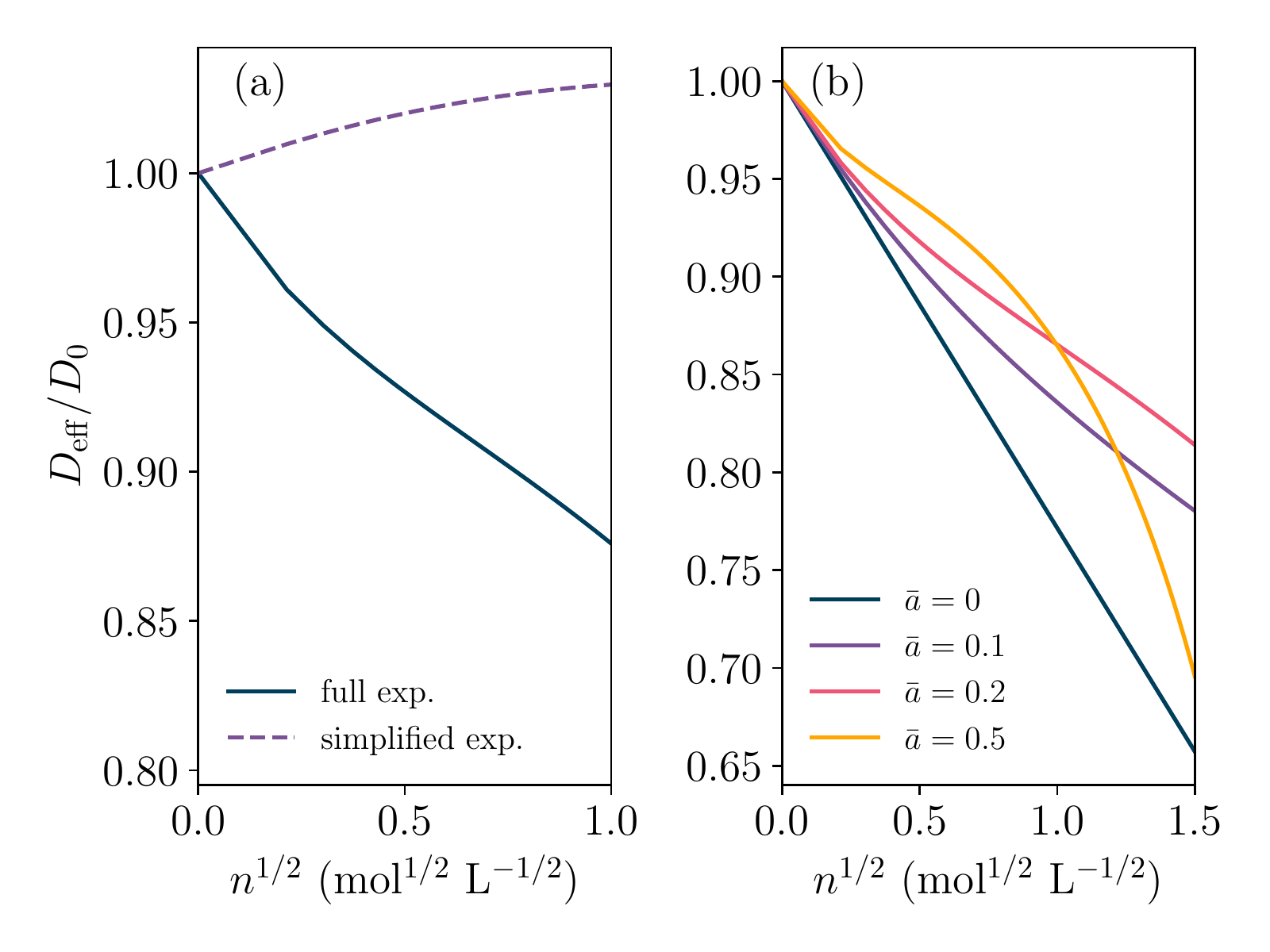}
\caption{\textbf{Self-diffusion: case where cations and anions have the same mobility.} (a) Self-diffusion coefficient as a function of the square-root concentration $n^{1/2}$ as obtained from the `full expression' [Eq.  \eqref{Deff_SDFT}] and the `simplified expression' [Eq. \eqref{Deff_SDFT_simp}]: the latter is obtained by neglecting the effect of the tracer on the bath. In both cases, we take $\bar{a}=a/\ell_B = 0.3$. (b) Self-diffusion coefficient as a function of the square-root concentration $n^{1/2}$ as given by Eq. \eqref{Deff_SDFT}, for several values of the rescaled cutoff value $\bar{a}$. For $\bar{a}=0$, we used the expression Eq. \eqref{Onsager}.}
\label{Deff_same_mobility}
\end{center}
\end{figure}

\subsection{Case where anions and cations have the same mobility}

We first focus on the particular case (which applies to KCl) where anions and cations have the same mobility. In this situation, we get the following expression for the effective diffusion coefficient of the tracer, which is very simple and explicit:
\begin{align}
\label{Deff_SDFT}
&\frac{D_\text{eff}}{D_0} = 1- \int_0^\infty \dd k\frac{8k^2}{3} {n \ell_B^2\cos^2(ka)} \nonumber \\
&\times \Big\{[4\pi \cos(ka)\ell_B n+k^2][8\pi \cos(ka)\ell_B n+k^2] \Big\}^{-1}.
\end{align}
This expression is plotted on Fig. \ref{Deff_same_mobility}(a): as expected, the self-diffusion coefficient of an ion in the electrolyte is a decreasing function of the concentration.

{A frequent simplification of the derivation presented in Section \ref{sec:path_integral} consists in neglecting the effect of the tracer on the dynamics of the density fields. This is referred to as the `passive case' in Ref. \citenum{Demery2011}), and has been used in different contexts\cite{Dean2007,Leitenberger2008,Marbach2018,wang2023interactions}.} Concretely, this is equivalent to neglecting the last term in the rhs of Eq. \eqref{field_mapping}. This yields the following expression for the effective diffusion coefficient of the tracer:
\begin{align}
\label{Deff_SDFT_simp}
&\frac{D_\text{eff}}{D_0} = 1+ \int_0^\infty \dd k\frac{32 \pi k^2}{3} {n^2 \ell_B^3\cos^3(ka)} \nonumber \\
&\times \Big\{[4\pi \cos(ka)\ell_B n+k^2]^2[8\pi \cos(ka)\ell_B n+k^2] \Big\}^{-1}
\end{align}
It is interesting to note that this simplified version actually predicts an unphysical behavior for the diffusion coefficient, which becomes an \emph{increasing} function of the overall concentration, and therefore exceeds its bare value as concentration increases (Fig. \ref{Deff_same_mobility}(a)). {This observation indicates that this simplification should be used with caution.}

\begin{figure*}
\begin{center}
\includegraphics[width=2\columnwidth]{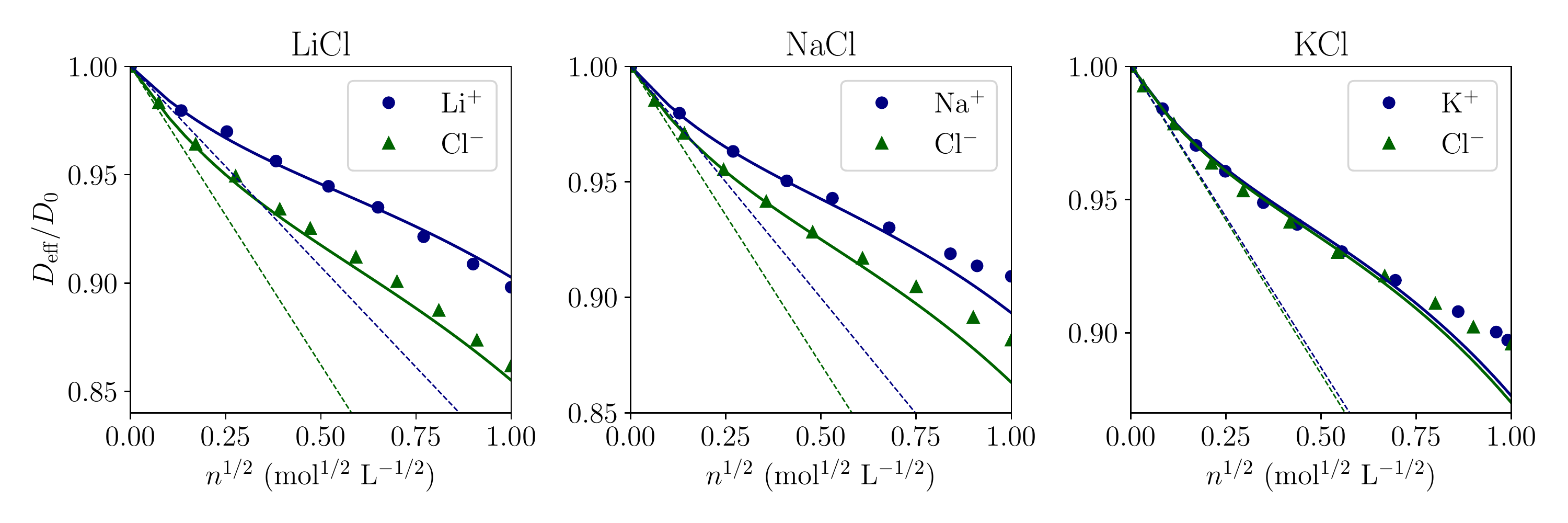}
\caption{{\bf Self-diffusion coefficients of ions in the electrolytes LiCl, NaCl and KCl:} results obtained from linearized SDFT (Eq. \eqref{Deff_full_exp}, solid lines) and from the MCT/MSA calculation (Ref.~\citenum{Dufreche2005}, symbols). Dashed lines: Onsager limiting law [Eq. \eqref{Onsager}].}
\label{compare_Dufreche}
\end{center}
\end{figure*}

\subsection{Case of a vanishing interaction radius ($a \to 0$)}

We now consider the limit of vanishing interaction radius ($a \to 0$) in Eq. \eqref{Deff_full_exp}. In this situation, the integral over $k$ can be computed exactly, and yields:
\begin{equation}
\lim_{a\to0}\frac{D_\text{eff}}{D_0} = 1-\frac{2\sqrt{\pi}}{3} \left( \sqrt{2}-\sqrt{\frac{1+3\Delta }{2(1+\Delta)}}\right) \sqrt{n \ell_B^3}.
\label{Onsager}
\end{equation}
Interestingly, this coincides exactly with the limiting law predicted by Onsager \cite{Onsager1945,Onsager1932}.

In order to investigate the accuracy of this simplified expression, we compare it to results obtained from Eq. \eqref{Deff_full_exp} with a fixed value of $\Delta$ and several values of the rescaled cutoff value $\bar{a}=a/\ell_B$ (Fig. \ref{Deff_same_mobility}(b)). Although the truncated potential is very simplified, we observe that introducing a cutoff value $a$ below which the potential vanishes has a significant impact on the self-diffusion coefficient of the tracer. A similar conclusion was reached when comparing the conductivity predicted by the Debye-H\"uckel-Onsager theory with the results from SDFT \cite{Avni2022}.

\subsection{General case and comparison to mode coupling theory/MSA}

We finally confront the result we obtained from linearized SDFT [Eq. \eqref{Deff_full_exp}] with earlier results obtained from a combination of mode-coupling theory (MCT) and the mean spherical approximation (MSA) \cite{Dufreche2002,Dufreche2005} . We consider three electrolytes: LiCl, NaCl and KCl. As input to our analytical theory, we use the values of the bare diffusion coefficients given in Ref. \citenum{Dufreche2002} and the values of the cutoff distances for the modified Coulomb potential given in Ref. \citenum{Avni2022}. These values are recalled in Table \ref{numerical_params}.
Results are shown on Fig. \ref{compare_Dufreche}. We find that, for these three electrolytes, the self-diffusion coefficients estimated from linearized SDFT calculations are very close to the MSA results. We may attribute this to the fact that the hydrodynamic contributions, that appear to be wrongly estimated by linearized SDFT (at least when it comes to the conductivity, see Section \ref{RPhydro}) do not have any effect on the self-diffusion coefficient at this order of approximation. A similar property held within the MSA/MCT treatment mentioned above.

\section{Importance of the choice of the modified potential}
\label{sec_mod_pot}

\begin{figure}
\begin{center}
\includegraphics[width=\columnwidth]{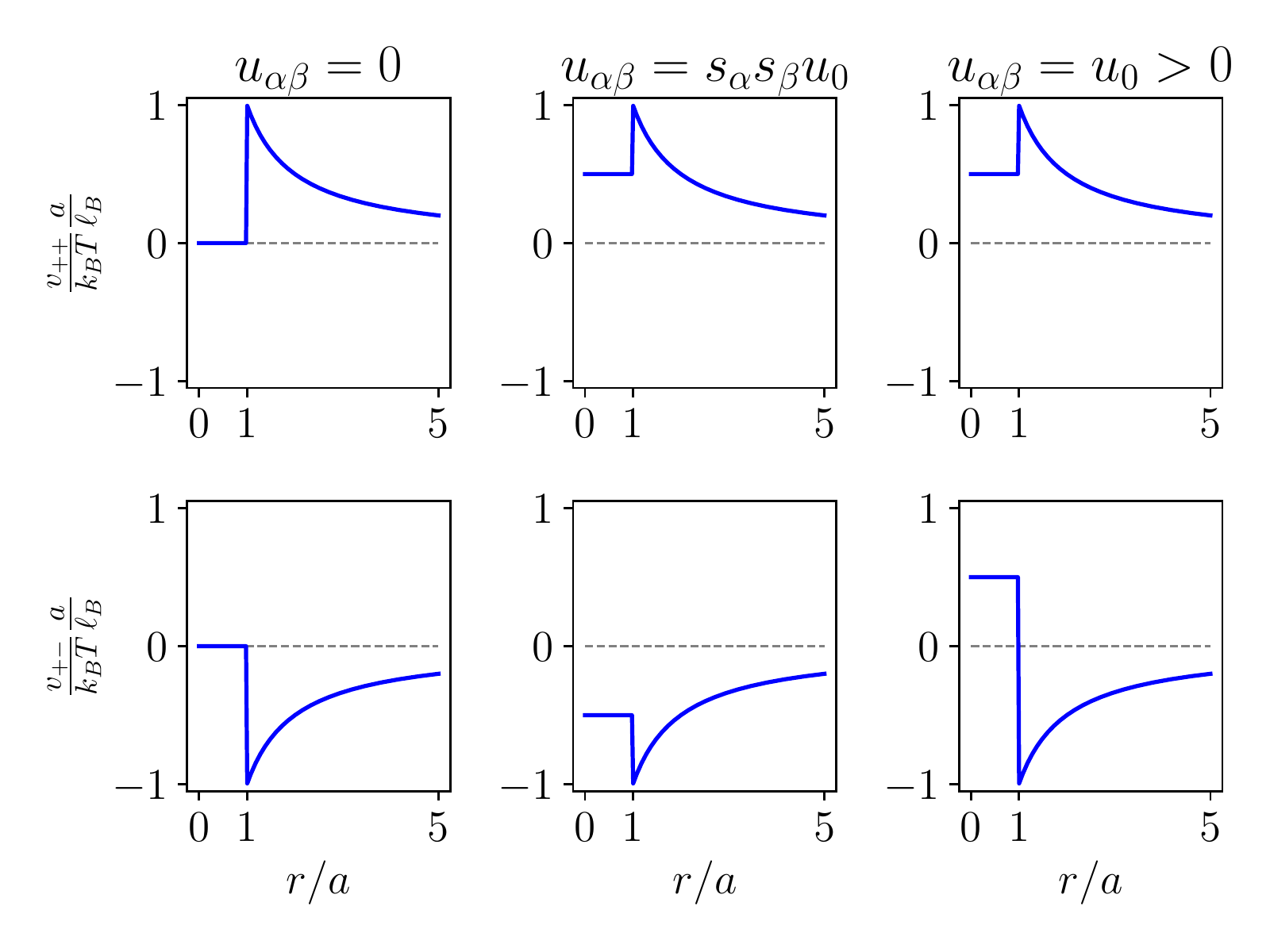}
\caption{\textbf{Modified interaction potentials:} as defined in Eq. \eqref{mod_pot}, for different choices of the parameter $u_{\alpha\beta}$. The last case, where $u_{\alpha\beta}=u_0$ for all pairs $(\alpha,\beta)$, is discussed in Appendix \ref{alternative_modif}.}
\label{potentials}
\end{center}
\end{figure}

\begin{figure*}
\begin{center}
\includegraphics[width=\columnwidth]{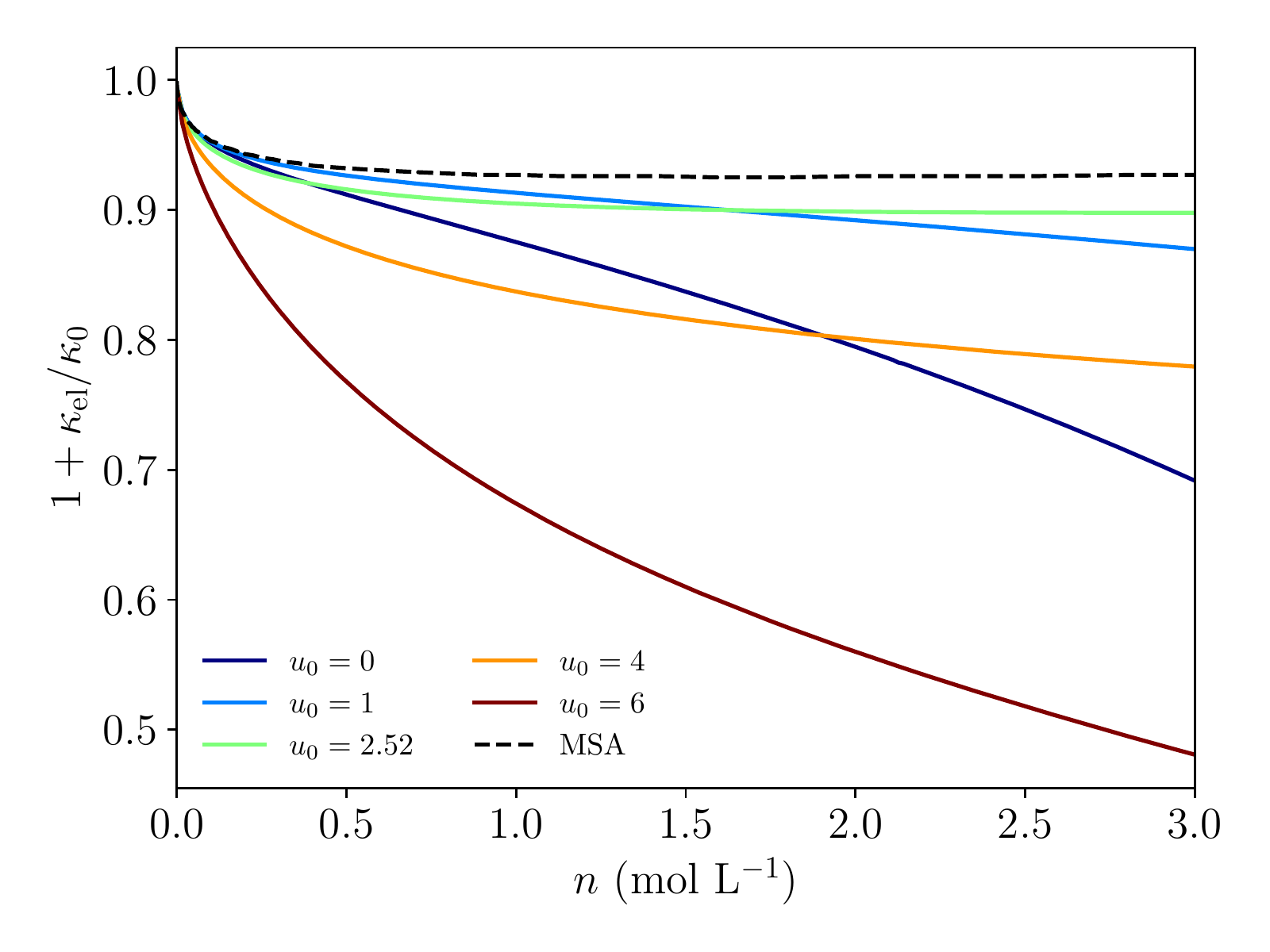}
\includegraphics[width=\columnwidth]{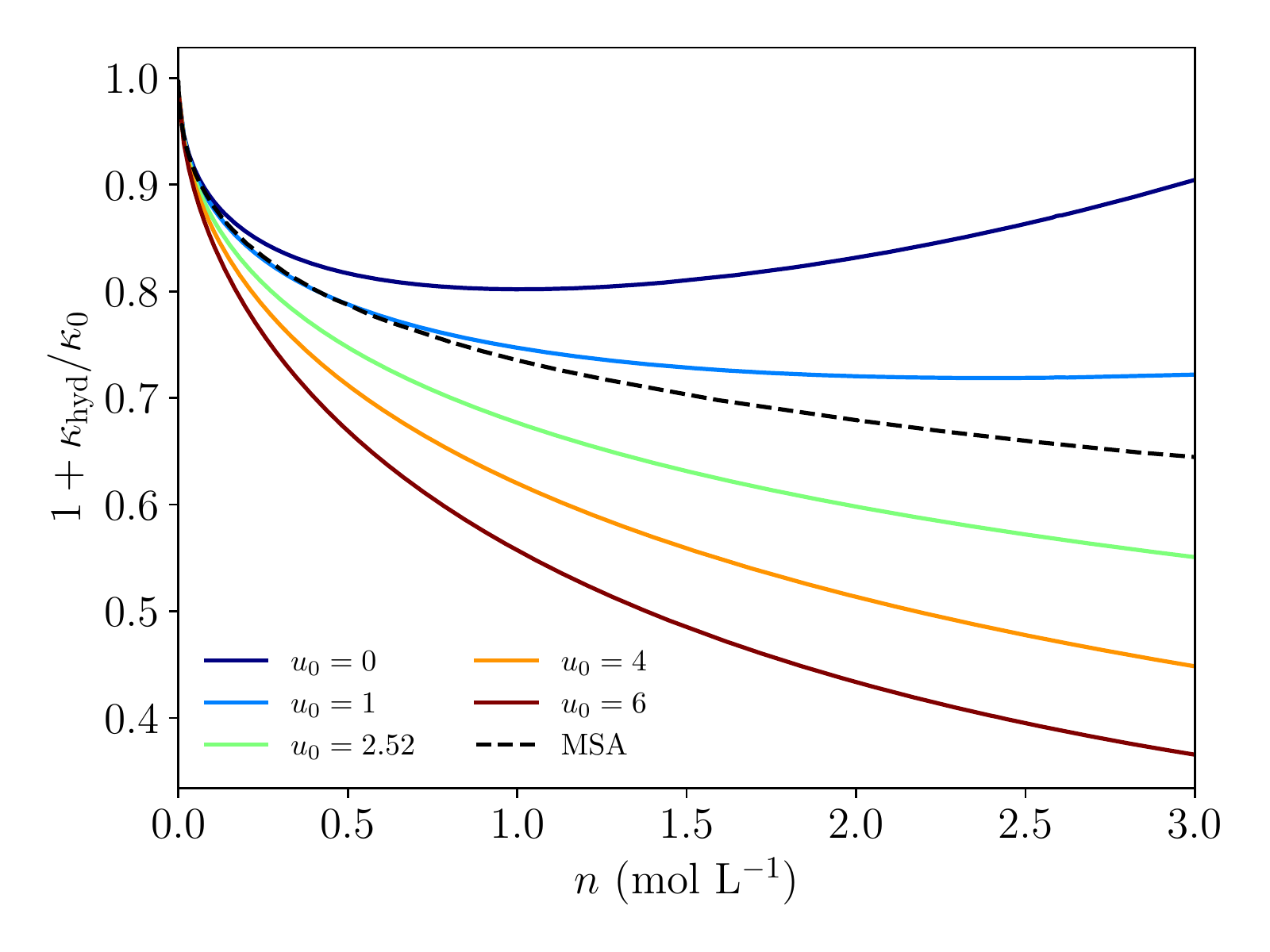}
\caption{\textbf{Influence of the repulsion parameter on the conductivity.} Electrostatic (left) and hydrodynamic (right) contributions to conductivity, with the modified Coulomb potential given in Eq.  \eqref{mod_pot}, and the repulsive contribution from Eq. \eqref{repulsive_option1}, for different values of the repulsion parameter $u_0$ (solid lines). The value $u_0=2.52$ corresponds to $u_0=a/\ell_B$. Dashed lines: results from MSA calculations [Eqs. \eqref{kappa_el_MSA} and \eqref{kappa_hyd_MSA}].}
\label{option2_influence_u0_conducti}
\end{center}
\end{figure*}

So far, we considered the modified Coulomb potential given in Eq. \eqref{mod_pot_Avni}, which is obtained by truncating the Coulomb interaction below a radius $r=a$ and setting the interaction energy to 0 below this cutoff.
 In this Section, we will show how the modifications of this potential, by changing the nature of the short-range part, can affect the observables we have considered so far, namely conductivity and self-diffusion. For this purpose, we rewrite the potential $v_{\alpha\beta}(r)$ under the form: 
 \begin{equation}
 \label{mod_pot}
v_{\alpha\beta}(r) = z_\alpha z_\beta V_\text{co}(r)+ u_{\alpha\beta} \theta(a-r).
\end{equation}
The first term corresponds to the truncated potential studied so far, the second term is a step `repulsion' term, written as a simple rectangular function  whose magnitude can be controlled with the parameter $ u_{\alpha\beta}$ ($ u_{\alpha\beta}=0$ corresponds to the case studied so far, and is represented on Fig. \ref{potentials}). In what follows, we study different possible choices for this parameter.

\subsection{Conductivity}

A first option consists in assuming that the  repulsion parameter $u_{\alpha\beta}$ depends on the considered pair. More precisely, we make the choice:
 \begin{equation}
 \label{repulsive_option1}
u_{\alpha\beta} = z_\alpha z_\beta u_0,
\end{equation}
in such a way that $u_{\alpha\beta} = u_0$ (resp. $-u_0$) if both ions have the same charge (resp. opposite charges). Although this is not the most physical choice (short-range repulsion is usually modeled by a charge-independent contribution), this choice is motivated by the fact that it can be used to make the interaction potential $v_{\alpha\beta}$ continuous at the cutoff value $r=a$, with the choice $u_0=\ell_B/a$. The case where the short-range contribution is charge-independent ($u_{\alpha\beta} = u_0$) is discussed in Appendix \ref{alternative_modif}.

 The Fourier transform of the modified potential now reads
\begin{equation}
\label{modified_potential_FT}
\tilde v_{\alpha\beta}(k) = \frac{\kB T}{2n\lambda_D^2 k^2 } \left[ \cos ka + \frac{u_0}{\ell_B k} (\sin ka -ka \cos ka) \right].
\end{equation}
It is straightforward to show that the corresponding expressions of $\kappa_\text{el}$ and $\kappa_\text{hyd}$ are obtained from Eqs. \eqref{kappa_el_Avni} and Eqs. \eqref{kappa_hyd_Avni} by making the substitution:
\begin{equation}
\cos\frac{ax}{\lambda_D} \to \cos\frac{ax}{\lambda_D} +\frac{\lambda_D u_0}{x \ell_B} \left( \sin\frac{ax}{\lambda_D} - \frac{ax}{\lambda_D} \cos\frac{ax}{\lambda_D}  \right)
\end{equation}
Importantly, the integrals involved in the expressions of $\kappa_\text{el}$ and $\kappa_\text{hyd}$ diverge in the limit $u_0\to\infty$, which is consistent with the general idea that linearized SDFT cannot account for hardcore interactions. We show on Fig. \ref{option2_influence_u0_conducti} the influence of the value of $u_0$ on the electrostatic contribution to the conductivity. Starting from $u_0=0$, which corresponds to the previous situation, and increasing $u_0$ up to  the value $(\ell_B/a)\kB T $, where the modified potential becomes continuous at $r=a$, we observe that the electrostatic contribution gets closer and closer to the value predicted by MSA. When $u_0 \gg \kB T$, the magnitude of the electrostatic contribution tends to diverge, as can be predicted from its analytical expression. Therefore, it seems that there exists an optimal value for the parameter $u_0$, that significantly improves the estimates from linearized SDFT when compared to earlier analytical schemes. A similar observation can be made about the hydrodynamic contribution, which appears to be closer to predictions from MSA calculations when $u_0$ is in the range $1-2~\kB T$ (Fig. \ref{option2_influence_u0_conducti}).

\begin{figure}
\begin{center}
\includegraphics[width=\columnwidth]{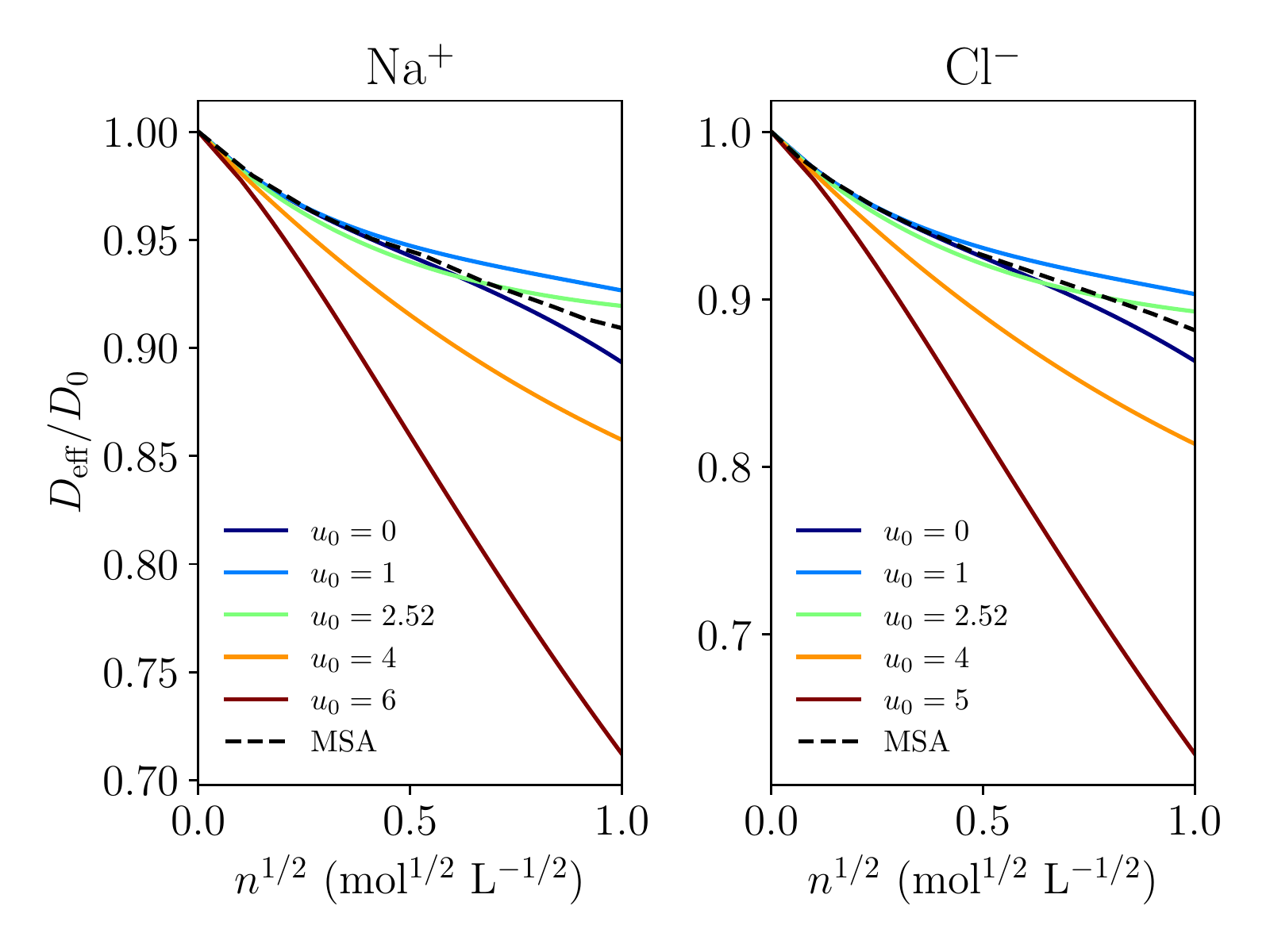}
\caption{\textbf{Influence of the repulsion parameter on self-diffusion.} Self-diffusion coefficients of ions in NaCl electrolyte, with the modified Coulomb potential given in Eq.  \eqref{mod_pot}, and the repulsive contribution from Eq. \eqref{repulsive_option1}, for different values of the repulsion parameter $u_0$ (solid lines). The value $u_0=2.52$ corresponds to $u_0=a/\ell_B$. Dashed lines: results from MSA/MCT calculations.}
\label{option2_influence_u0_self_diff}
\end{center}
\end{figure}

\subsection{Self-diffusion}

The effect of the value of the repulsion parameter on self-diffusion can be studied in a similar way, by modifying the results obtained in Section \ref{sec_self_diff}. Technically, in Eq. \eqref{Deff_full_exp}, we replace the potential $\tilde u$ by the expression given in Eq. \eqref{modified_potential_FT}. We observe that, just like the conductivity, the self-diffusion coefficient is highly sensitive to the value of the repulsion parameter, and that its choice significantly affect the quality of the predictions from linearized SDFT when compared to results from the MCT-MSA framework (Fig. \ref{option2_influence_u0_self_diff}).

\section{Conclusion}

In this work, we discussed several analytical theories for the conductivity and self-diffusion in concentrated electrolytes. Among them, linearized SDFT has been used quite extensively during the past years, and it allowed very successful predictions in the limit where  electrolytes are very dilute. Recently, this analytical method was extended to study concentrated electrolytes, in which the short-range repulsion between ions plays a predominant role. A small-distance truncation of the Coulomb potential, which is tractable within linearized SDFT, has been put forward in order to account for short-range effects in a very simple fashion.

We showed that, within this approximation, the output of linearized SDFT contradicts earlier analytical (Mean-Spherical Approximation) and numerical (Brownian dynamics) estimates of the conductivity of concentrated electrolytes. However, the linearized SDFT treatment can be improved in several ways: (i) the hydrodynamic effects can be accounted at the Rotne-Prager level rather than at the Oseen tensor, which regularizes the hydrodynamic contribution to conductivity, or which at least shifts any unphysical divergence to higher densities; {(ii) the short-range interaction energy between ions, which was set to zero in earlier treatments, can instead be set to a finite value, which may indeed improve the validity of the results by Avni et al.\cite{Avni2022} when compared to other analytical schemes such as MSA.} Finally, we also computed the self-diffusion coefficient of ions within SDFT: as opposed to the conductivity, it appears that the dependence of this observable over ionic concentrations can be captured more easily by the truncated interaction, but that it is still highly dependent on the short-range details of the potential.

In conclusion,  linearized SDFT is a particularly appealing tool to compute transport and diffusion properties in fluctuating systems of interacting particles, given its relative simplicity and ability to yield explicit expressions. However, we emphasize that, in spite of its advantages, it should be handled with caution, in particular when it comes to describing effects dominated by the short-range interactions between particles. Earlier analytical schemes, which often treat these interactions in a more thorough manner, should be used as a guide for the approximations that are implemented within linearized SDFT. Finally, we emphasize that computational approaches that rely on an implicit or coarse-grained representation of the solvent, and which often correspond to the same level of description than SDFT, constitute a very fertile field of research\cite{Ladiges2022,Kuron2016,Tischler2022}, that usefully complement analytical descriptions.

\begin{acknowledgments}
The authors thank David Andelman, Ram Adar, Henri Orland, Yael Avni, Aleksandar Donev and Jean-Fran\c{c}ois Dufr\^eche  for discussions. 
This project received funding from the European Research Council under the European Union's Horizon 2020 research and innovation program (project SENSES, grant Agreement No. 863473).
\end{acknowledgments}

\appendix

\section{Direct correlation functions under the Mean Spherical Approximation}
\label{c_MSA}

In this Appendix, we recall the expression of the direct correlation functions that can be derived under the Mean Spherical Approximation (see Section \ref{app_MSA_outline} for details and references). These functions can be written under the form:
\begin{equation}
    \tilde c_{\alpha\beta}(k)=-\frac{4\pi z_\alpha z_\beta\ell_B}{k^2}+{\tilde c}^s_{\alpha\beta}(k).
\end{equation}
While the random phase approximation consists in assuming ${\tilde c}^s_{\alpha\beta}(k)=0$, the mean spherical approximation yields:
\begin{align}
    {\tilde c}^s_{\alpha\beta}(k) = & \frac{4\pi a^3}{K^6}[24 d_{\ab}-2 b_{\ab} K^2+e_{\ab}K^4\nonumber\\
    &-[24 d_{\ab} -2(b_{\ab}+6d_{\ab})K^2 \nonumber\\
    &+(a_{\ab}+b_{\ab}+d_{\ab}+e_{\ab})K^4] \cos K \nonumber\\
   & +[-24 d_{\ab} K +(a_{\ab}+2b_{\ab}+4d_{\ab})K^3] \sin K],
\end{align}
where we use the shorthand notation $K=ka$ for the dimensionless wavevector, and where we define:
\begin{align}
    a_{\alpha\beta}&=-\frac{(1+2\eta)^2}{(1-\eta)^4}-2B\left (\frac{a}{\lambda_D} \right)\frac{\ell_B}{a} z_\alpha z_\beta, \\
 b_{\alpha\beta}    &=-\frac{6\eta(1+\eta/2)^2}{(1-\eta)^4}+\left[B\left( \frac{a}{\lambda_D} \right)\right]^2 \frac{\ell_B}{a} z_\alpha z_\beta , \\
     d_{\alpha\beta}&= -\frac{\eta(1+2\eta)^2}{2(1-\eta)^4}, \\
     e_{\ab} &= \frac{\ell_B}{a}z_\alpha z_\beta, \\
     B(x)&=\frac{x^2+x-x\sqrt{1+2x}}{x^2},
\end{align}
with $\eta=\frac{\pi}{6}a^3(n_+ + n_-)$ the total packing fraction.

\section{Alternative modification of the Coulomb potential}
\label{alternative_modif}

In this Appendix, we consider an alternative modification of the Coulomb potential, and assume that the repulsive part does not depend on the charge of the ion ($u_{\alpha\beta}=u_0$ for all pairs $\alpha$, $\beta$). We actually consider the following general expression for the pair potential:
 \begin{equation}
\label{mod_pot_appendix}
v_{\alpha\beta}(r) = z_\alpha z_\beta \frac{e^2}{4\pi \varepsilon_0\varepsilon r}\theta(r-a) + v_\text{rep}(r),
\end{equation}
where the repulsive part can first remained unspecified.
 With this choice, the symmetry relation $V_{++}=-V_{+-}$ does not hold anymore. Consequently, the field-field correlations (which are found as the solutions of the linearized DK equations for the cation and anion density fields) have different symmetries than in the situations considered before. First, we find  that, under these conditions, the repulsive part of the potential $v_\text{rep}$ has no influence on the hydrodynamic contribution. Therefore it is sill given by Eq. \eqref{kappa_hyd_Avni}.

\begin{figure}
\begin{center}
\includegraphics[width=\columnwidth]{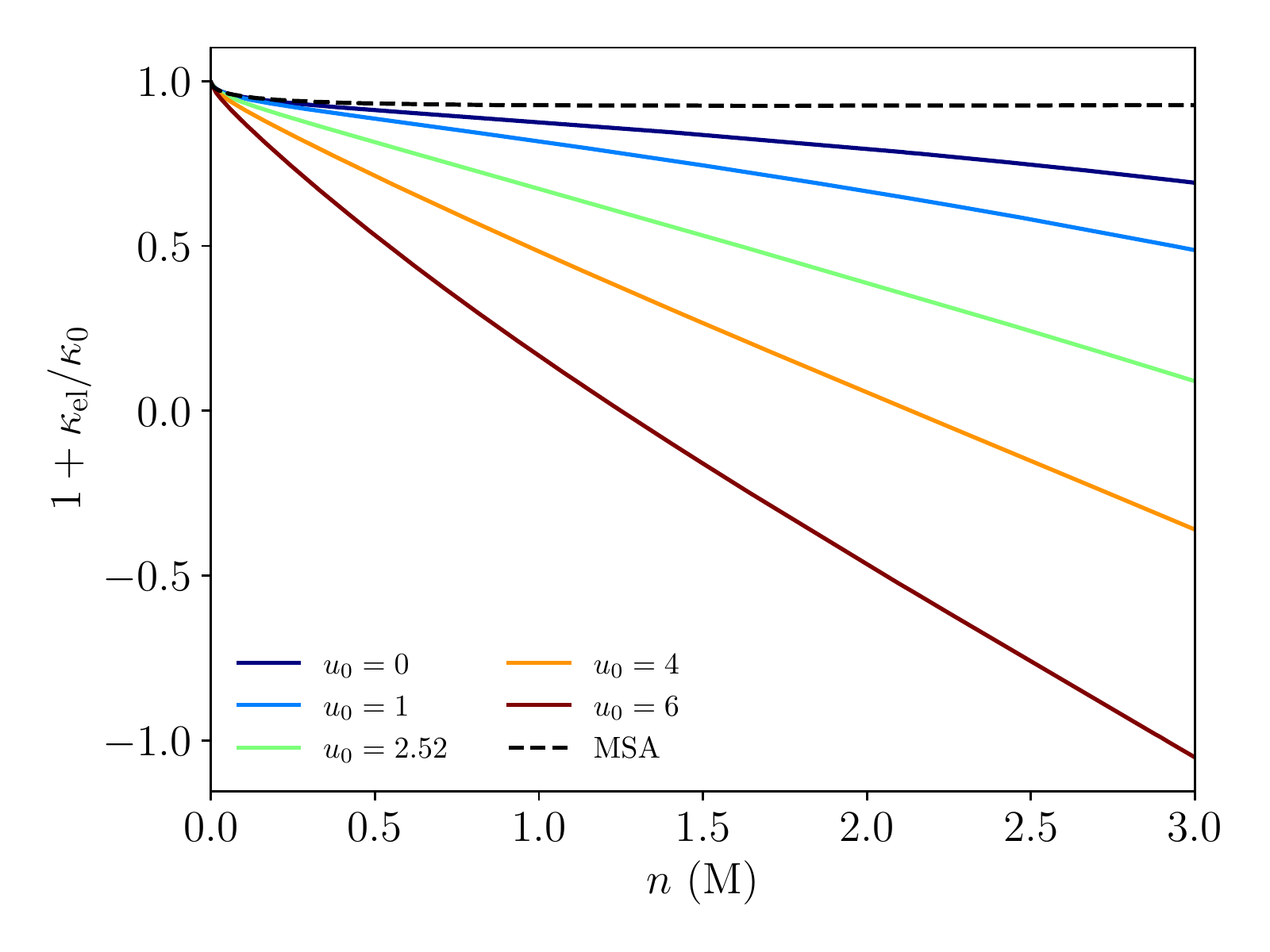}
\caption{\textbf{Influence of the repulsion parameter on the conductivity.} Electrostatic contribution to conductivity, with the modified Coulomb potential given in Eq.  \eqref{mod_pot_appendix} for different values of the repulsion parameter $u_0$ (solid lines). The value $u_0=2.52$ corresponds to $u_0=a/\ell_B$.  Dashed line: results from MSA [Eq. \eqref{kappa_el_MSA}].}
\label{fig_appendix}
\end{center}
\end{figure}

Second, we find the following expression for the electrostatic contribution: 
  \begin{align}
\label{ }
&\kappa_\text{el} = -2e \bar{\mu} \int \frac{\dd \kk}{(2\pi)^3} {n^2 k_z^2 (\tilde{v}_\text{co}(k)-\tilde{v}_\text{rep}(k))^2}\nonumber\\
&\times \left[4 k^2 (\frac{1}{2}+n \tilde{v}_\text{co}(k)) (\frac{1}{2}+n \tilde{v}_\text{rep})[1+n(\tilde{v}_\text{co}+\tilde{v}_\text{rep})]\right]^{-1}
\end{align}
  where we introduced $\tilde{v}_\text{co}(k) = \tilde{V}_\text{co}(k)/\kB T$ and $\tilde{v}_\text{rep}(k) = \tilde{V}_\text{rep}(k)/\kB T$. In order to get a more explicit expression, we write the repulsive part as before, under the form $V_\text{rep}(r) = \kB T u_0 \theta(a-r)$.  We get the following expression for the electrostatic contribution: 
\begin{widetext}
\begin{equation}
\label{ }
\kappa_\text{el}=-\frac{1}{3\pi} \frac{\kappa_0 \ell_B}{\lambda_D} \int_0^\infty \dd x \frac{x^2\left(\cos\frac{ax}{\lambda_D}-\frac{u_0\lambda_D}{\ell_B x}\varphi  \left( \frac{ax}{\lambda_D}    \right)\right)^2}{\left( x^2+\cos\frac{ax}{\lambda_D} \right)\left( 1+\frac{u_0\lambda_D}{\ell_B x^3}\varphi\left( \frac{ax}{\lambda_D}    \right) \right)\left( x^2+\frac{1}{2}\cos\frac{ax}{\lambda_D}+\frac{u_0\lambda_D}{2 \ell_B x}\varphi\left( \frac{ax}{\lambda_D}    \right) \right)},
\end{equation}
\end{widetext}
where we introduced the shorthand notation $\varphi(X)=\sin X- X\cos X$. We show on Fig. \ref{fig_appendix} the electrostatic contribution to conductivity as a function of the electrolyte concentration, for different values of the parameters $u_0$. As opposed to the case considered in Section \ref{sec_mod_pot}, it is difficult to improve the predictions from  linearized SDFT with this choice of the repulsive contribution.

%\bibliography{sdft_electrolytes}


\end{document}